%% file: main.tex
\documentclass[acmsmall,screen]{acmart}
\usepackage{graphicx} 

\makeatletter
\def\@ACM@checkaffil{
    \if@ACM@instpresent\else
    \ClassWarningNoLine{\@classname}{No institution present for an affiliation}%
    \fi
    \if@ACM@citypresent\else
    \ClassWarningNoLine{\@classname}{No city present for an affiliation}%
    \fi
    \if@ACM@countrypresent\else
        \ClassWarningNoLine{\@classname}{No country present for an affiliation}%
    \fi
}
\makeatother

\input{gabe_macros}

\definecolor{mygreen}{HTML}{467E7E}
\definecolor{mygray}{rgb}{0.5,0.5,0.5}
\definecolor{myblue}{HTML}{144B7D}
\definecolor{myorange}{HTML}{B25A00}
\lstdefinestyle{myC}{
  language=C,
  backgroundcolor=\color{white},
  basicstyle=\linespread{0.9}\ttfamily\tiny,
  breakatwhitespace=false,
  breaklines=true,
  commentstyle=\bfseries\color{red},    
  deletekeywords={...},
  escapeinside={<@}{@>},
  extendedchars=true,
  keepspaces=true,
  keywordstyle=\bfseries\color{myblue!70}, 
  otherkeywords={uint},     
  deletekeywords={get,angle, gamma, invariant},
  emph = { certiq_prove, match, end},
  emphstyle=\bfseries\color{myorange!70},
  showspaces=false, 
  showstringspaces=false,
  showtabs=false, 
  stringstyle=\color{mymauve},
  tabsize=1,
 moredelim=**[is][{\btHL[fill=red!40]}]{`}{`},
}

\newsavebox{\measurebox}

\newcommand{\approach}[0] {SymPrompt\xspace}
\newcommand{\evalfms}[0] {897\xspace}

\newcommand{\gabe}[1] {\textcolor{pistachio}{(Gabe: #1)}\xspace}
\newcommand{\bray}[1] {\textcolor{mred}{(Baishakhi: #1)}\xspace}
\definecolor{pastel-red}{HTML}{FF9AA2}
\newcommand{\sid}[1] {\textcolor{pastel-red}{(SJ: #1)}\xspace}

\newcommand{\todo}[1] {\textcolor{red}{(TODO: #1)}\xspace}

\renewcommand{\todo}[1]{}
\renewcommand{\gabe}[1]{}
\renewcommand{\sid}[1]{}
\renewcommand{\bray}[1]{}

\setcopyright{rightsretained}
\acmDOI{10.1145/3643769}
\acmYear{2024}
\copyrightyear{2024}
\acmSubmissionID{fse24main-p940-p}
\acmJournal{PACMSE}
\acmVolume{1}
\acmNumber{FSE}
\acmArticle{43}
\acmMonth{7}
\received{2023-09-29}
\received[accepted]{2024-01-23}

\author{Gabriel Ryan}
\authornote{Work done while the author was an intern at AWS AI Labs}
\email{gabe@cs.columbia.edu}
\affiliation{\institution{Columbia University} \country{USA}}
\author{Siddhartha Jain}
\authornote{Corresponding author}
\email{siddjin@amazon.com}
\author{Mingyue Shang}
\email{myshang@amazon.com}
\author{Shiqi Wang}
\email{wshiqi@amazon.com}
\author{Xiaofei Ma}
\email{xiaofeim@amazon.com}
\author{Murali Krishna Ramanathan}
\email{mkraman@amazon.com}
\author{Baishakhi Ray}
\email{rabaisha@amazon.com}
\affiliation{\institution{AWS AI Labs} \country{USA}}

\renewcommand{\shortauthors}{{Ryan, Jain, Shang, Wang, Ma, Ramanathan, and Ray}}

\begin{abstract}
\input{sections/00_abstract}
\end{abstract}



\title{Code-Aware Prompting: A Study of Coverage-Guided Test Generation in Regression Setting using LLM}

\begin{document}
{
\sloppy
\maketitle
}
\input{sections/01_introduction}

\input{sections/02_motivation}
\input{sections/03_approach}
\input{sections/05_evaluation}

\input{sections/06_threats_to_validity}
\input{sections/07_discussion}
\input{sections/08_related_work}

\input{sections/09_conclusion}


\bibliographystyle{ACM-Reference-Format}
\bibliography{main}


\end{document}

%% file: gabe_macros.tex
\usepackage{comment}

\usepackage{booktabs} 
\usepackage{xspace}
\usepackage{longtable}
\usepackage{graphicx}
\usepackage{algorithm}
\usepackage[noend]{algpseudocode}
\usepackage{listings}
\usepackage{url}

\usepackage{tabularx}

\usepackage{amsfonts}
\usepackage{amsthm}
\usepackage{multirow}
\usepackage{amsmath}
\usepackage[english]{babel}

\newcommand{\eg}[1]{(e.g., #1)\xspace}
\newcommand{\ie}[1]{(i.e., #1)\xspace}
\usepackage{xcolor}
\usepackage{subcaption}
\usepackage[inline]{enumitem}
\usepackage{longfbox}

\usepackage{array}
\newcolumntype{L}[1]{>{\raggedright\let\newline\\\arraybackslash\hspace{0pt}}m{#1}}
\newcolumntype{C}[1]{>{\centering\let\newline\\\arraybackslash\hspace{0pt}}m{#1}}
\newcolumntype{R}[1]{>{\raggedleft\let\newline\\\arraybackslash\hspace{0pt}}b{#1}}

\usepackage[font=small]{caption}

\definecolor{mred}{rgb}{.80,.12,.30}
\definecolor{grey}{rgb}{0.5,0.5,0.5}
\definecolor{purple2}{rgb}{.75,0,.85}
\definecolor{pistachio}{rgb}{0.35, 0.60, 0.20}
\definecolor{torange}{HTML}{ff7f0e}
\definecolor{steelblue}{rgb}{.10,.15,.9}

\usepackage{tikz}

\makeatletter
\newenvironment{btHighlight}[1][]
{\begingroup\tikzset{bt@Highlight@par/.style={#1}}\begin{lrbox}{\@tempboxa}}
	{\end{lrbox}\bt@HL@box[bt@Highlight@par]{\@tempboxa}\endgroup}

\newcommand\btHL[1][]{%
	\begin{btHighlight}[#1]\bgroup\aftergroup\bt@HL@endenv%
	}
	\def\bt@HL@endenv{%
	\end{btHighlight}%
	\egroup
}
\newcommand{\bt@HL@box}[2][]{%
	\tikz[#1]{%
		\pgfpathrectangle{\pgfpoint{1pt}{0pt}}{\pgfpoint{\wd #2}{\ht #2}}%
		\pgfusepath{use as bounding box}%
		\node[anchor=base west, fill=orange!25,outer sep=.5pt,inner xsep=0.5pt, inner ysep=-0.3pt, rounded corners=2pt, minimum height=\ht\strutbox-.1pt,#1]{\raisebox{.01pt}{\strut}\strut\usebox{#2}};
	}%
}
\makeatother

\definecolor{lred}{rgb}{.95, .3, .3}
\definecolor{lgreen}{rgb}{.3, .65, .3}

\newcommand{\ttt}[1]  {\texttt{#1}}

\newcommand{\tbf}[1]  {{\bf #1}}
\newcommand{\circled}[1]{\raisebox{.5pt}{\textcircled{\raisebox{-.9pt} {#1}}}}

\newcommand{\parheader}[1] {\noindent\textbf{#1}\xspace}



%% file: sections/00_abstract.tex
Testing plays a pivotal role in ensuring software quality, yet conventional Search Based Software Testing (SBST) methods often struggle with complex software units, achieving suboptimal test coverage. Recent work using large language models (LLMs) for test generation have focused on improving generation quality through optimizing the test generation context and correcting errors in model outputs, but use fixed prompting strategies that prompt the model to generate tests without additional guidance. As a result LLM-generated testsuites still suffer from low coverage.

In this paper, we present SymPrompt, a code-aware prompting strategy for LLMs in test generation. SymPrompt's approach is based on recent work that demonstrates LLMs can solve more complex logical problems when prompted to reason about the problem in a multi-step fashion. We apply this methodology to test generation by deconstructing the testsuite generation process into a multi-stage sequence, each of which is driven by a specific prompt aligned with the execution paths of the method under test, and exposing relevant type and dependency focal context to the model. Our approach enables pretrained LLMs to generate more complete test cases without any additional training. We implement SymPrompt using the TreeSitter parsing framework and evaluate on a benchmark challenging methods from open source Python projects. SymPrompt enhances correct test generations by a factor of 5 and bolsters relative coverage by 26\% for CodeGen2. Notably, when applied to GPT-4, SymPrompt improves coverage by over $2\times$ compared to baseline prompting strategies. 

%% file: sections/01_introduction.tex
\section{Introduction}

Testing is an essential component of software development that allows developers to catch bugs early in the development lifecycle and prevents costly releases of buggy software~\cite{planning2002economic}. 
However, manual test writing can be time-consuming, taking up more than 15\% of development time on average~\cite{daka2014survey}. 
Extensive research has therefore been devoted to developing automated test generation approaches, which can automate the process of writing testsuites for software units under development.
Automated test generation for the purposes of generating a testsuite that becomes part of development codebase is typically performed in a \emph{regression setting}, which assumes that the code currently under test is implemented correctly, and the objective is to generate a suite of tests that will effectively detect \emph{future} bugs that may be introduced in to the codebase during later development, causing a regression~\cite{yoo2012regression}.

\paragraph{Limitation of Existing Approaches.} Widely used test-suite generation tools such as Evosuite~\cite{fraser2011evosuite} use a Search Based Software Testing (SBST) approach in which test inputs are randomly generated and mutated to maximize coverage of the software unit under test. However, SBST approaches struggle to generate high coverage test inputs in many cases, such as when branch conditions depend on specific values or states that are difficult to resolve with randomized inputs and heuristics. In a large scale study of SBST on 110 widely used open source projects, Fraser et al. observed that more than 25\% of the tested software classes had less than 20\% coverage~\cite{fraser2014large}.

The limitations of SBST have motivated recent work using large language models (LLMs) for automated testsuite generation~\cite{tufano2020unit}. 
In this setting, LLMs are typically instructed to generate test cases by supplying the source code of the focal method and optionally some additional code context. The instruction with which as user interacts with an LLM is commonly referred to as {\em prompt}. 

Unlike SBST approaches that reason about the execution behavior of a method under test (commonly called {\em focal method}), LLMs approximate the overall functionalities of the focal method based on 
the natural naming convention~\cite{allamanis2015suggesting, ahmad2020transformer} of its implementation (e.g., meaningful method name, variable names, etc.), API usage, and calling context. 
This capability of LLMs can be harnessed to create test cases, specifically generating inputs targeting branch conditions that necessitate particular input values or states. 
However, when tasked with generating test inputs for methods with challenging  or complex branch conditions, 
LLMs usually succeed in producing inputs for the easy branches, leaving 
behind the branches with more complex conditions. 
Consequently, for hard real-world use cases, LLMs usually struggle to generate high coverage testsuites~\cite{dinella2022toga, alagarsamy2023a3test}.

Figure \ref{fig:working_ex} illustrates how a focal method can pose challenges for both SBST and LLM-based test generation approaches. The method takes a string representing a filesystem object as input and categorizes the type of filesystem object based on an external method call. Generating high coverage test inputs for this method is very difficult for a SBST approach because generating strings that represent different filesystem devices is extremely unlikely with randomized input generation, and in practice it will only test paths for which it has preprogrammed heuristics to generate string inputs such as directories (see Figure \ref{fig:working_ex:pynguin}). Conversely, an LLM trained on a large code corpus should in theory have the knowledge to generate strings representing filesystem objects like block devices and sockets, but in practice, when given an open ended prompt to implement a set of tests for the method, will only generate tests for relatively simple and common inputs for a filesystem utility function, such as temporary files and directories (see Figure \ref{fig:working_ex:llm_baseline}).

\paragraph{Our Solution.} In this work we introduce a novel approach, \approach, to constructing prompts 
leveraging different code properties, which enables LLMs to generate test inputs for more complex focal methods. 
Recent research employing LLMs for logical problem-solving demonstrates that when LLMs are prompted to decompose the problem into multiple stages of reasoning first instead of directly trying to generate the answer, they exhibit greatly enhanced capability in solving more complex problems~\cite{wei2022chain, wei2022emergent, yao2023tree}. We build on this concept devising a unit test specific decomposition framework and integrating it into a novel multi-stage prompting strategy.
Our key insight is that the process of generating a test-suite to fully cover a focal method can be broken down into a sequence of logical 
problems that are posed to the LLM prompts.
For each prompt, the LLM will generate an appropriate test case, and thus, will eventually generate a series of test cases with higher code coverage to test the diverse behavior of the focal method. 

At a high level, \approach works in three stages: (i) Given a focal method, \approach tries to statically capture the execution behavior of the method 
by collecting approximate path constraints and return values for each execution path. (ii) \approach collects some properties of the focal method including argument types, external library dependencies, code context, etc. (iii) For each execution path, \approach constructs a prompt using the collected information and solicits the LLM to generate a test input that will execute the corresponding path. This generation is carried out iteratively, and the test cases generated in previous iterations are appended to subsequent prompts to provide further guidance to the LLM in generating a thorough testsuite.

Conceptually, our approach resembles symbolic analysis based test generation techniques. However, traditional symbolic analysis is known to 
suffer in real-world code with complex data types, external dependencies, and complex branching behavior. The proposed work addresses these  limitations in three ways.
First, \approach 
collects approximate constraints based on static code rather than attempting to resolve all data types and unresolved dependencies while collecting path constraints.
For instance, instead of attempting to symbolically reason about a call like \texttt{os.is\_dir()}, we keep it as it is and rely on the LLM to reason about the underlying functionality of the method call based on its name and usage.
Second, to mitigate computational overhead arising from numerous branching conditions, we focus only on the paths with unique branching conditions. Finally, instead of relying on a solver to resolve underlying constraints, we employ an LLM to generate test cases based on the prompt. The insight here is that, although the LLM may not precisely reason about all possible constraints, it typically extracts sufficient information from the code context and approximate path constraints to generate meaningful test inputs.

Our step-by-step prompting approach further facilitates the test generation process. Without this type of prompting, the LLM would need to (1) infer what branch the previous test it generated covered, (2) search for a new branch to address, and (3) create a test for it. In our approach, we delegate steps 1 and 2 to traditional static analysis, allowing the LLM to focus solely on step 3.

\paragraph{Results.} 
We implement \approach for Python using the TreeSitter parsing framework and evaluate on a benchmark of \evalfms focal methods that are challenging for existing SBST test generation. We prototype our technique on open source CodeGen2 16B LLM: \approach improves the ratio of correct test generations by a factor of 5 and improve relative coverage by 26\%.
To check generalizability of our technique, we further evaluate \approach with a state-of-the-art LLM, GPT-4: \approach improves relative coverage by 105\% over tests generated with a baseline prompting strategy.

In summary, this paper makes the following contributions:
\begin{enumerate}
    \item We introduce \emph{Path Constraint Prompting}, a novel, code-aware prompting strategy for LLM test generation that breaks the process of generating a testsuite into a multi-stage procedure of generating test inputs for each execution path in the method under test.
    \item We implement our approach in \approach using the TreeSitter parsing framework with integrations for both open-source transformers and GPT models. \approach is currently a proprietary research prototype and we are working with our legal team to make the code publicly available. 
    \item We evaluate \approach on a benchmark of \evalfms methods that are challenging for existing SBST approaches from widely used Python projects and show it improves relative coverage by 26\% for CodeGen2 and by a factor of more than $2\times$ (105\% relative improvement) when used with GPT-4.  
\end{enumerate}

%% file: sections/02_motivation.tex
\begin{figure}
\centering
\sbox{\measurebox}{%
  \begin{minipage}[b]{.43\textwidth}
    \begin{subfigure}[b]{\linewidth}
        \includegraphics[width=\textwidth]{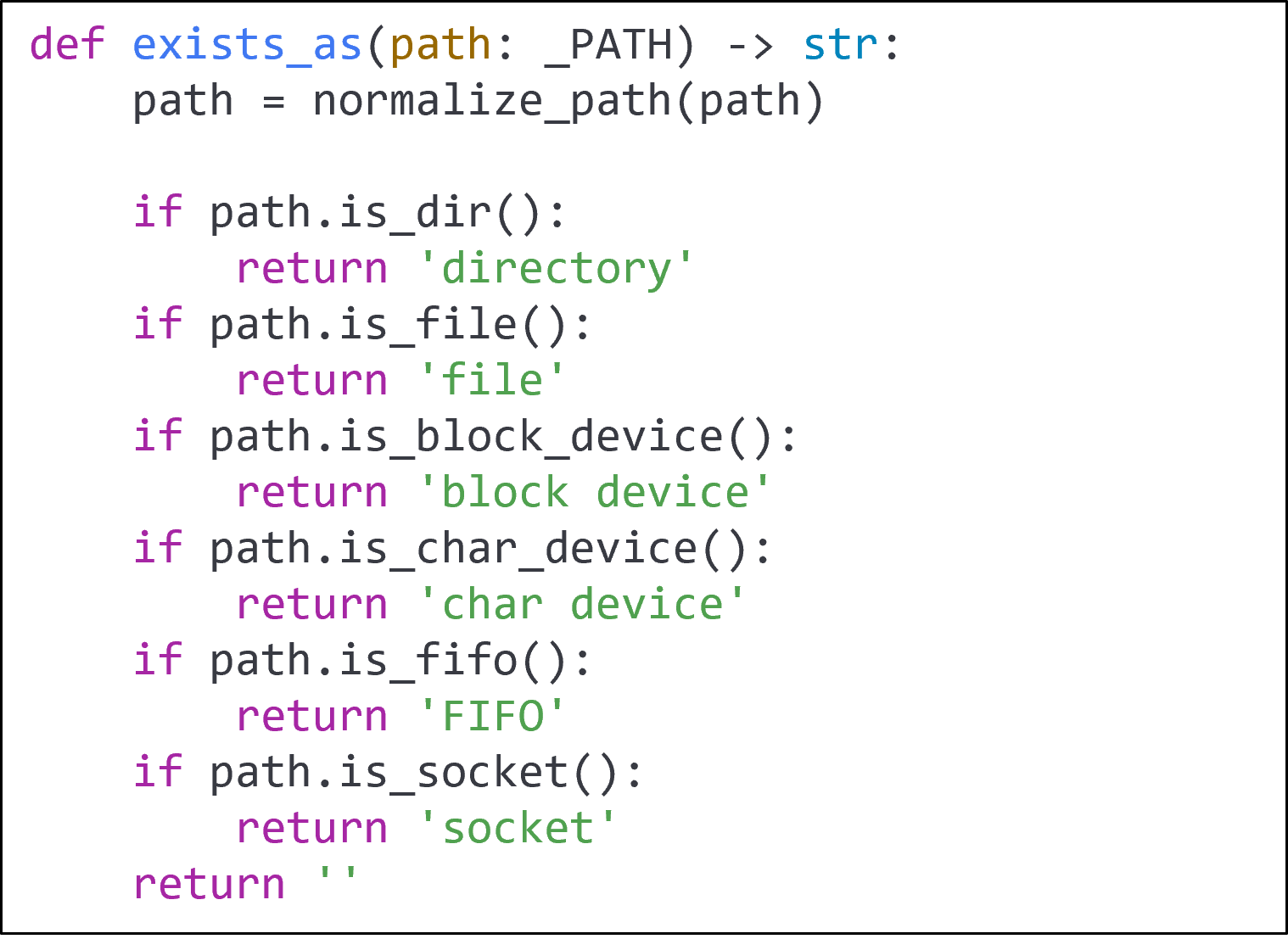}
        \vspace{-10pt}
        \caption{Focal method.\label{fig:working_ex_fm}}
    \end{subfigure}
    
    \vspace{25pt}
    
    \begin{subfigure}[b]{\linewidth}
        \includegraphics[width=\textwidth]{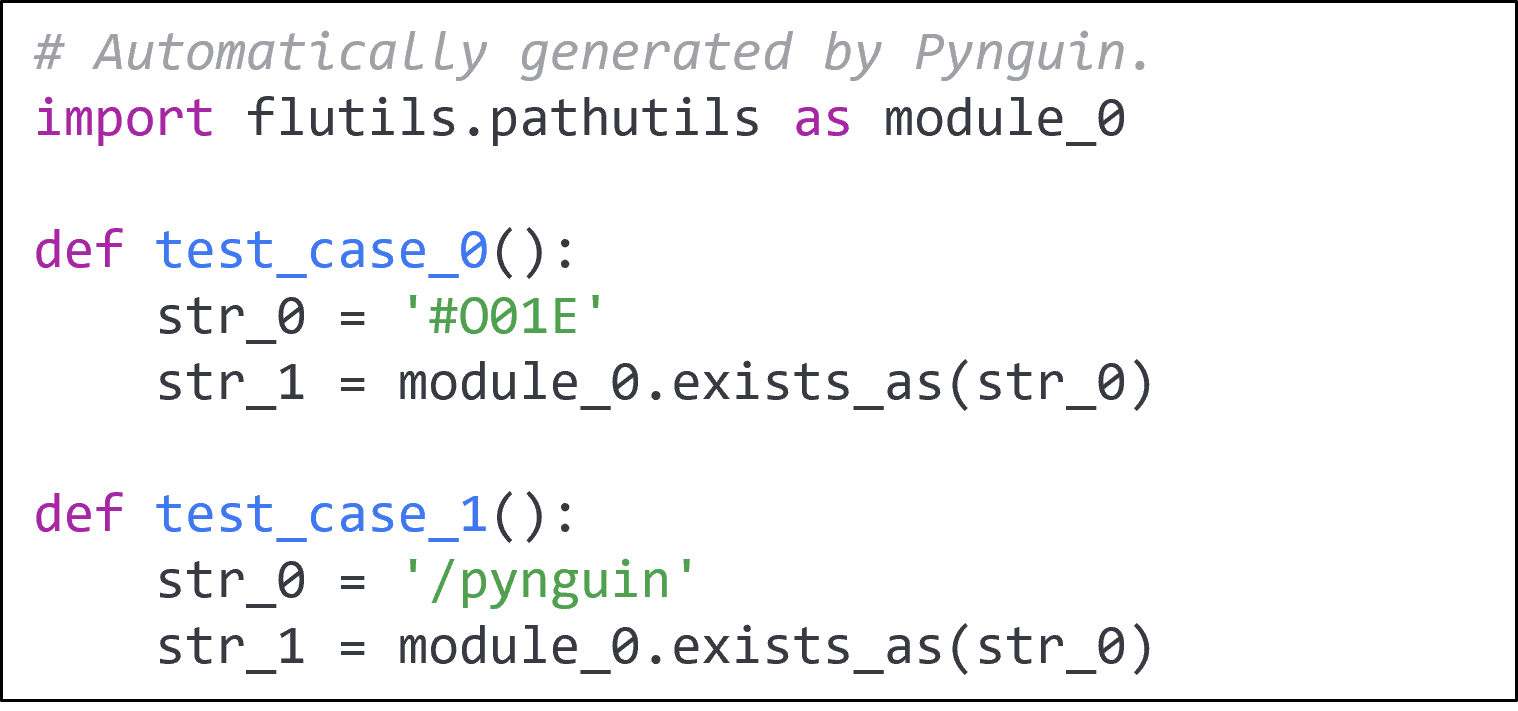}
        \vspace{-10pt}
        \caption{SBST test generations.\label{fig:working_ex:pynguin}}
    \end{subfigure}
  \end{minipage}
}
\usebox{\measurebox}\qquad
\begin{minipage}[b][\ht\measurebox][s]{.43\textwidth}
\centering
  \begin{subfigure}[b]{\linewidth}
        \includegraphics[width=\textwidth]{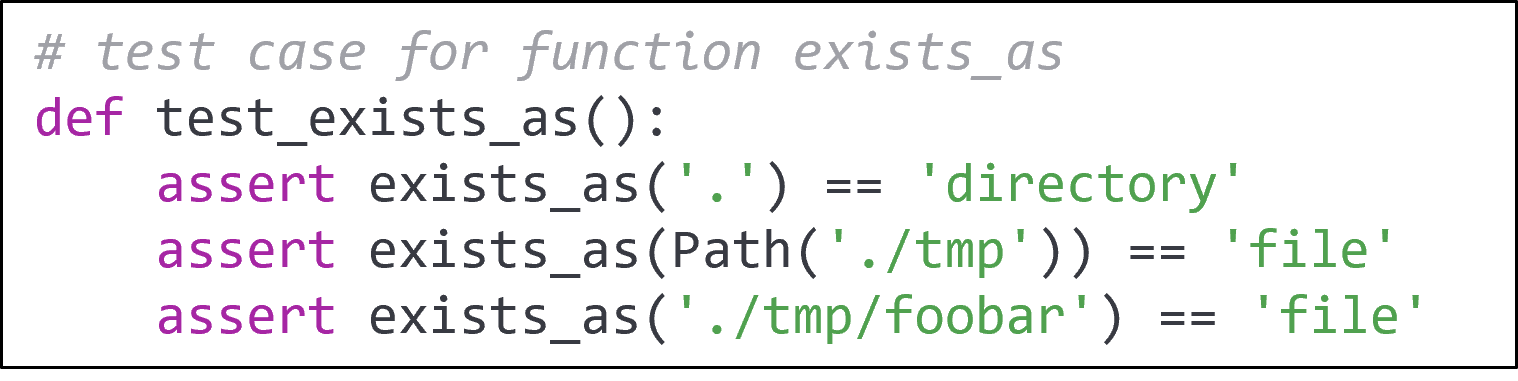}
        \vspace{-10pt}
        \caption{LLM test generations.\label{fig:working_ex:llm_baseline}}
    \end{subfigure}

    \vfill
    
    \begin{subfigure}[b]{\linewidth}
        \includegraphics[width=1.\textwidth]{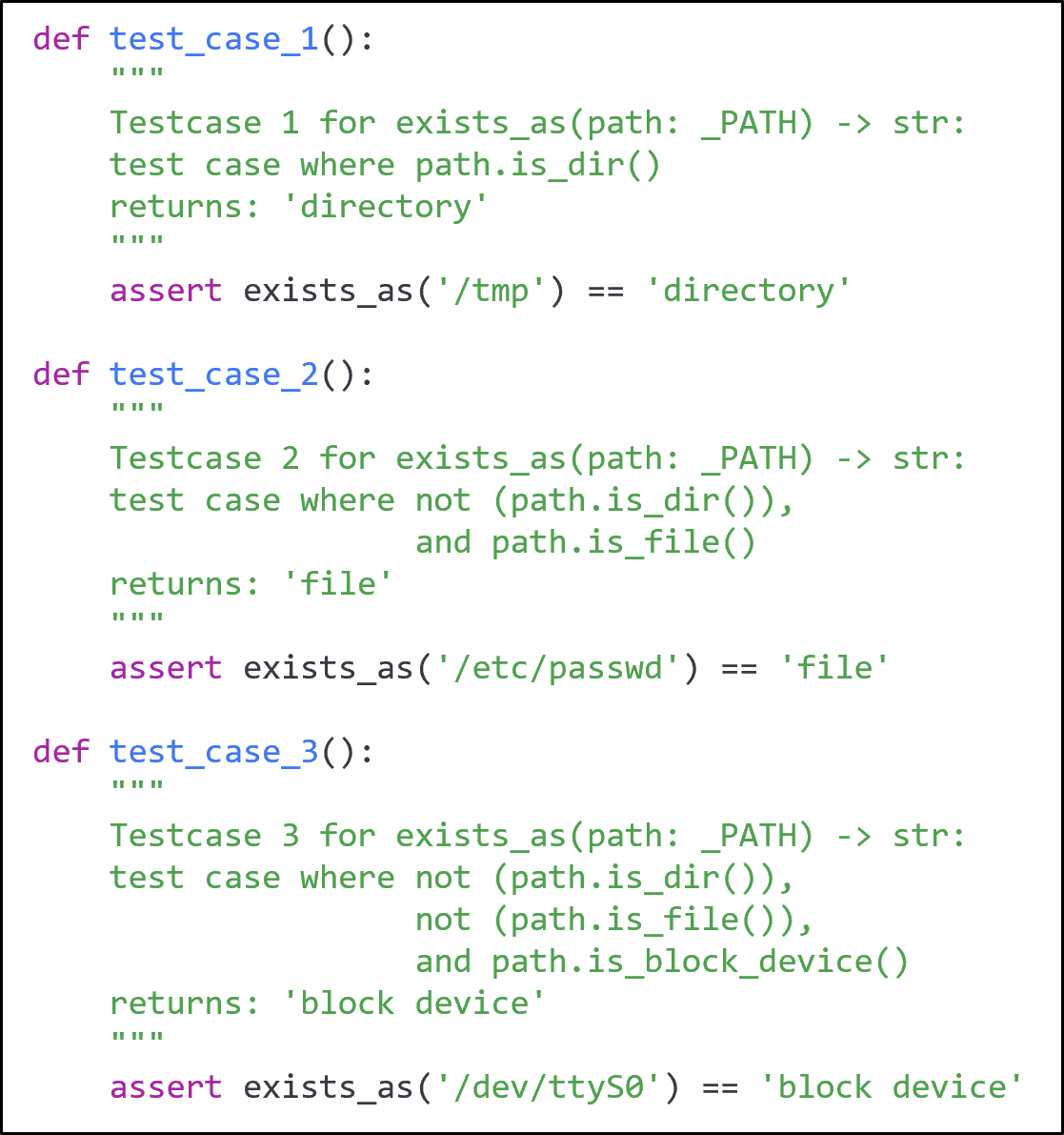}
        \caption{\approach test generations.\label{fig:working_ex:symprompt}}
\end{subfigure}
\end{minipage}

\caption{Example test generations from an SBST tool (Pynguin), zero shot LLM (CodeGen2), and \approach prompts for focal method \ttt{exists\_as} in the \ttt{flutils} open source Python project. An SBST approach is unable to generate full coverage tests for this method without special configuration because it is unable to generate strings that represent specific types of filesystem objects \eg{block devices}. An LLM conversely is able to generate input strings associated with filesystem objects such as block devices, but in practice will only test a small subset of use cases based on the most likely usage scenarios such as paths to files and directories. \approach constructs path specific prompts to guide the model to generate high coverage testsuites.\label{fig:working_ex}}
\end{figure}

\begin{figure}
\centering
\includegraphics[width=0.95\textwidth]{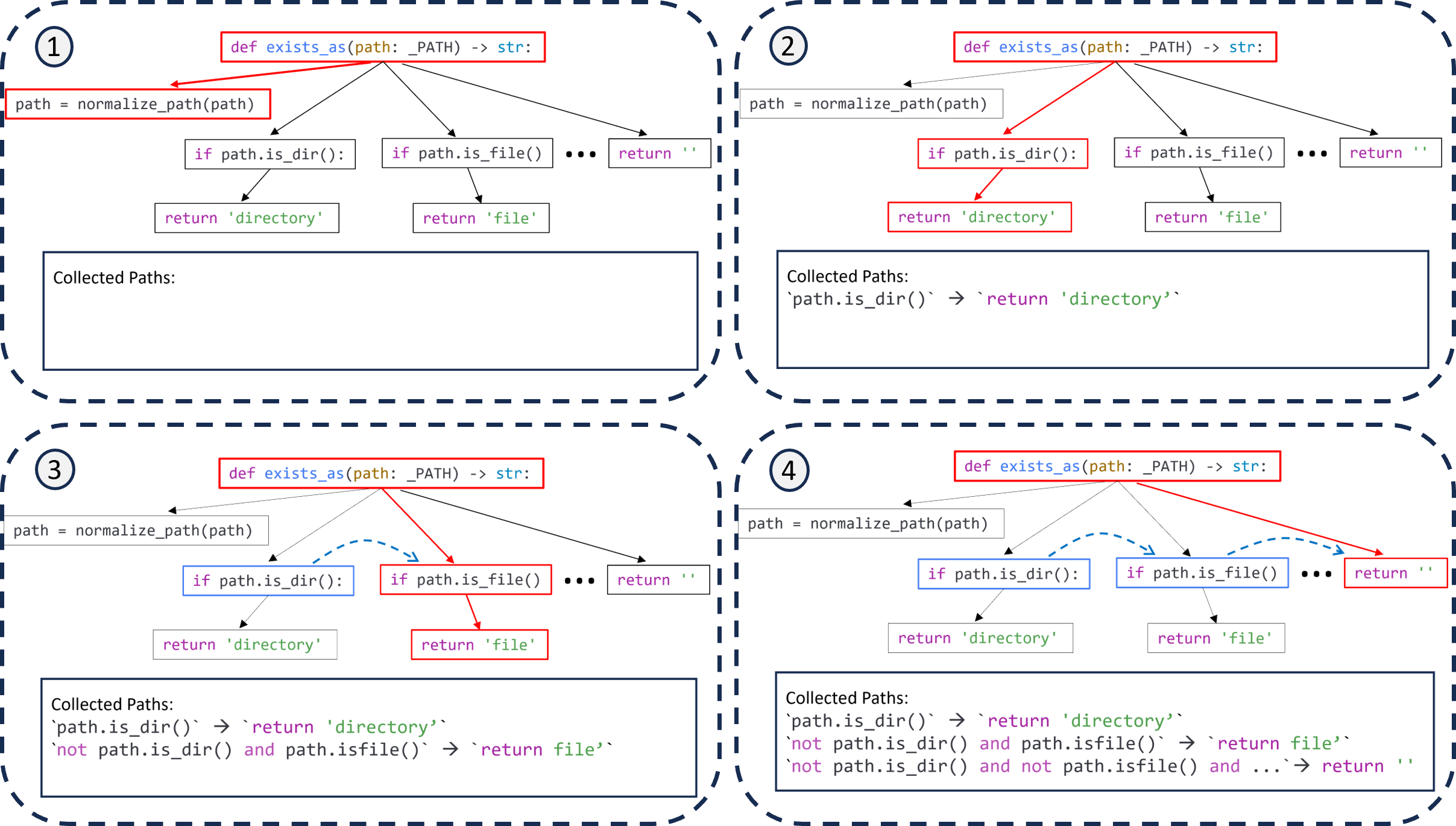}
\caption{Workflow for generating path constraint prompts. The focal method shown in Figure \ref{fig:working_ex_fm} is first parsed and its abstract syntax tree is traversed in preorder. In step \circled{1}, the traversal first visits the first method statement, \ttt{normalize\_path(path)}, but does not record any information since it is not a branch constraint. In step \circled{2}, it then traverses to the first \ttt{if} statement, and records that there is a constraint \ttt{path.is\_dir()} that must be satisfied to execute the current path on the AST. It then reaches the \ttt{return 'directory'} under the first \ttt{if} check, and records that there is an execution path where 'directory' is returned when \ttt{path.is\_dir()} is true. The  preorder traversal next visits the \ttt{if path.is\_file()}, \ttt{return 'file'} branch of the AST in \circled{3} and records a second path where \ttt{path.is\_dir()} is \ttt{false} and \ttt{path.is\_file()} is \ttt{true}, and the return behavior is \ttt{'file'}. This traveral continues until in step \circled{4}, the final \ttt{return} statement is reached, based on an execution path where none of the branch constraints are \ttt{true}. Each collected execution path and return value is then used to construct  prompts for test generations that specifies both the path constraints and return behavior for the target test case.
}
\label{fig:workflow}
\end{figure}

\section{Working Example}
\label{sec:working_ex}


In this section we provide a working example of how path constraint prompts are constructed and used to guide an LLM generate high coverage tests in a regression setting. 
Figure \ref{fig:working_ex_fm} illustrates a focal method, named \ttt{exists\_as}, extracted from our evaluation within the \texttt{flutils.path\_utils} module. Testing this method poses a significant challenge for both SBST (Search-Based Software Testing) and LLM-based approaches due to its extensive branching structure, which requires tailored input data to achieve full coverage of test cases. Each branch in this method demands specific input values that satisfy precise constraints---particular string inputs that reference filesystem objects like directories and block devices. 

\parheader{SBST-based Test Generation.} Figure \ref{fig:working_ex:pynguin} shows a suite of regression tests generated with the Python SBST tool \texttt{Pynguin}. Since SBST approaches generate inputs randomly according pre-programmed heuristics, they are unlikely to generate inputs that represent specific filesystem objects such as block devices and sockets unless the tool was specifically configured to generate strings representing these objects as inputs. In the case of Pynguin, the input strings it generated in Figure \ref{fig:working_ex:pynguin} represent a nonexistent device (\texttt{`\#O01E'}) and a directory that is defined in the Pynguin test environment (\texttt{`/pynguin'}) that cover two of return behaviors in the focal method, but do not test the other use cases that require other specific input values. Pynguin is not capable of generating input strings that test these other use cases without special configuration. In addition, it generates tests that do not follow common usage patterns in developer written tests, which makes the tests more difficult to maintain~\cite{tufano2020unit}.

\gabe{two additional things to potentially discuss here. (1) a common sbst strategy is to chain API calls and feed return values from one API call into inputs in the next. However, this does not help in this use case because \ttt{flutils.pathutils} does not include APIs for obtaining filesystem object name strings. (2) SBST/symbolic testing methods often include some symbolic analysis to obtain input values from branch comparison values. However, this cannot be applied to this function because the branch constraints are based on another method call \ttt{normalize\_path} that returns a \ttt{Path} object which in turn has the methods that are checked in each branch condition \eg{\ttt{is\_dir()}}. So both API call chaining and simple symbolic constraint solving do not work for this function.}

\parheader{LLM-based Test Generation.} Figure \ref{fig:working_ex:llm_baseline} shows a set of regression test calls generated using a standard code-completion testing prompt used by Lemieux et al.~\cite{lemieux2023codamosa}. Unlike an SBST tool such as Pynguin, an LLM has an 
approximate domain understanding to reason that a branch constraint like \ttt{path.is\_file()} will likely be tested by an input like \ttt{`./tmp'} and can generate associated input strings to test these use cases, even without observing the definitions of \ttt{normalize\_path} and \ttt{is\_file} on the first and second lines of code in \ttt{exists\_as} in Figure \ref{fig:working_ex_fm}. However, in cases where focal methods have many different paths, and some paths represent less common use cases \eg{if the input string is a block device}, LLMs will only generate test cases for the most common use cases of a focal method, even if many test generations are sampled. In this case, the LLM (CodeGen2) only generates tests for the two most common uses, where the input is either a directory or a file.

\vspace{0.2cm}
\parheader{\approach.} Our approach works in three steps:
(i) \textit{Collecting Approximate Path Constraints.} This step is the core to our approach. Figure \ref{fig:workflow} shows how path constraints are collected and used to prompt an LLM to generate high coverage tests. Possible execution paths in the focal method are collected by traversing the method's abstract syntax tree (AST) in preorder and recording branch conditions on each possible execution path, where if a branch is not taken its branch condition is inverted on a given path. When a return statement is reached, the set of branch constraints that need to be satisfied to reach that return statement is recorded, along with the returned value. Each recorded path constraint and return value is then used to generate a  prompt, which instructs the LLM to generate a test case targeting the corresponding path. 

\begin{figure}
\centering
\includegraphics[width=0.9\textwidth]{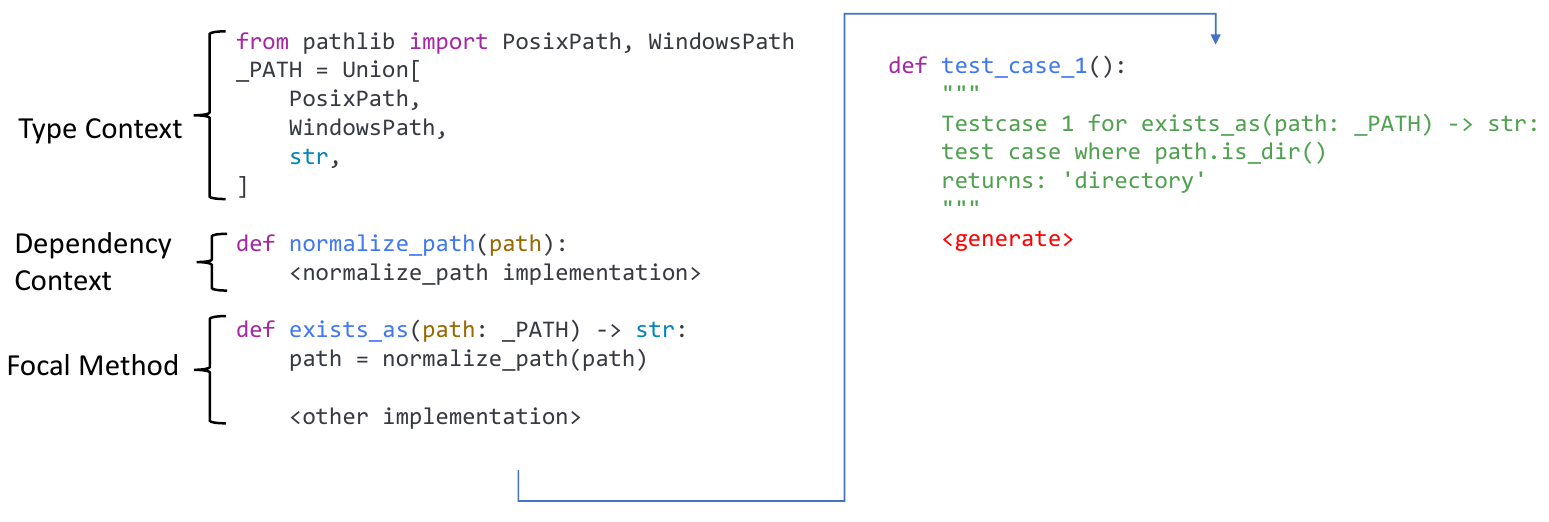}
\caption{Example of generation used by \approach. The prompt exposes both the type and dependency context of the focal method to the model in addition to the path constraint prompt. \label{fig:context_ex}}
\end{figure}

(ii) \textit{Context Construction.} 
Besides the path constraint, each prompt includes the focal method signature, which guides the LLM to generate a correct focal method call. We further include additional focal context for generation based on the types and methods that appear in the focal method that exposes relevant data structure and external library dependencies the model may need to reference for effective test generation, as shown in Figure \ref{fig:context_ex}. For the focal method \ttt{exists\_as}, we include the definition of the input parameter \ttt{\_PATH} and the external method call \ttt{normalize\_path}. One such prompt for our working example is shown in Figure \ref{fig:context_ex}.


(iii)~\textit{Test Generation.} For the focal method \ttt{exists\_as}, path constraint prompting guides the LLM to generate tests for use cases it does not normally cover without specific prompting. In particular, path constraint prompts guide the model to generate test strings that satisfy the constraints for \ttt{is\_block\_device()}, \ttt{is\_char\_device()}, \ttt{is\_fifo()}, and \ttt{is\_socket()}. When specifically prompted, CodeGen2 is able to generate correct test inputs in three of these four cases in our evaluation, more than doubling the number of tested branches covered in the generated testsuite (see Figure~\ref{fig:working_ex:symprompt}).

Path constraint prompting, in conjunction with type-aware focal contexts, allows us to leverage advantages of LLMs in deriving meaningful test inputs from context, understanding how to correctly initialize complex input types, utilizing relevant external API calls in testing, and resolving difficult-to-cover branch constraints, while deriving the advantages of a coverage-aware testing strategy similar to SBST by prompting for a set of high coverage tests.



\todo{background CoT, incontext learning, make sure to cite the android bug replay paper~\cite{feng2023prompting} and any similar works using CoT or incontext in a similar way }

%% file: sections/03_approach.tex
\section{Methodology}

In this section, we elaborate on our approach, \approach, to generating test cases for a given focal method in a regression setting.
At the core of our approach is crafting tailored prompts
that break the problem of test generation into multiple stages of reasoning based on the possible execution paths in focal method. These prompts are designed to instruct the model to generate test cases for a specific set of execution paths within the focal method that will ensure comprehensive branch and line coverage.

\subsection{Overview}


Our approach to constructing multi-stage reasoning prompts is based on static symbolic analysis techniques~\cite{godefroid2007compositional, king1976symbolic, baldoni2018survey} to reason about underlying program behavior.  
At a high level, for each control path ($\rho$) of a focal method, 
we statically collect a path constraint $\Phi_\rho$
that will steer the program execution along $\rho$.

Once the path constraints are collected, in the conventional symbolic analysis-based testing approach, a systematic search algorithm is employed to enumerate all the path constraints~\cite{godefroid2005dart}. 
Feasible paths are those for which the corresponding $\Phi_\rho$ is satisfiable---any concrete solution that meets the conditions of $\Phi_\rho$ can be used to execute and test the path $\rho$ in a regression setting.

However, when dealing with complex real-world code (for instance, when input parameters of the focal method are of complex data types, or when focal methods rely on external dependencies and API calls), the traditional symbolic analysis-based techniques often encounters difficulty in finding a solution. Furthermore, in cases where focal methods feature numerous nested branches and conditions, symbolic execution frequently experiences significant computational overhead.

In this work, we overcome the above mentioned shortcomings in three ways.
First, instead of trying to resolve all the data types and unresolved dependencies while collecting path constraints, we allow approximations. 
For example, if the method \texttt{path.is\_dir()} in Figure~\ref{fig:workflow} cannot be reasoned about symbolically, we leave the condition as it is while collecting $\phi_\rho$. 
Second, to address computational overhead with many branching conditions, we only focus on the paths having unique branching conditions. This design decision significantly helps to reduce the number of paths.
Finally, instead of using a solver to solve the underlying constraints we leverage an LLM to generate test cases based on the collected path constraints. Our insight is that, although the LLM may not reason precisely about all possible constraints, it usually derives enough information from code context and the approximate path constraints 
to generate meaningful test inputs.



\begin{figure}
\centering
        \includegraphics[width=0.9\textwidth]{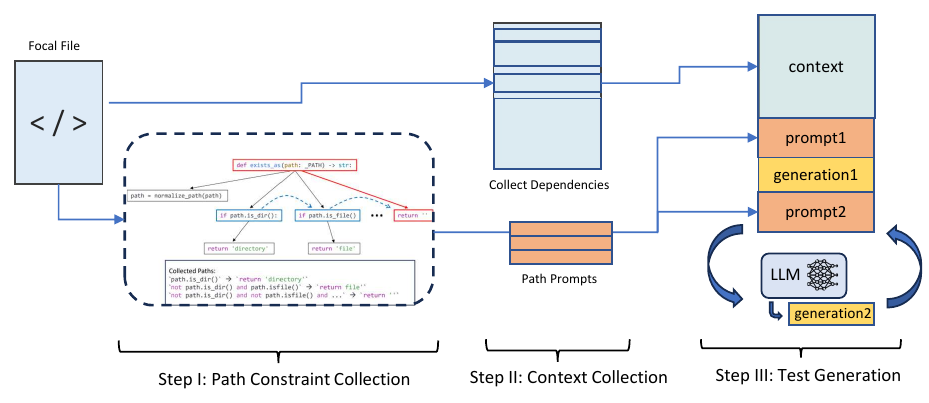}
\caption{Overview of \approach's framework for test generation. 
In Step-I, path constraint collection is performed on the focal method. In Step-II, the type and dependency context from the focal method are parsed from the focal file along with the focal method itself. Finally, in Step-III, prompts for each set of path constraints are then constructed and iteratively passed to the model to generate test cases.\label{fig:end2end}}
\end{figure}

To this end, our test generation contains three steps: 
(Step-I) collecting approximate path constraints and return expressions for program paths, 
(Step-II) capturing relevant code context, and
(Step-III) generating prompts amenable to a LLM by leveraging the path constrains and context identified in the above step. This entire process is done statically without executing the focal method.
Figure \ref{fig:end2end} gives a high level overview of how we use \approach to generate tests.

\subsection{Step-I: Collecting Approximate Path Constraints}


The objective of \approach is to generate test inputs that will follow specific execution paths in the focal method. To facilitate this, we construct prompts to expose the model to relevant path constraints. 
This step describes in details how we collect approximate path constraints by statically traversing the abstract parse tree (AST) of the focal method. 

\subsubsection{Path Constraint Collection} 

Our approach to collecting path constraints is similar to static symbolic execution. We perform a preorder traversal of the focal method abstract parse tree and collect constraints for each branch and loop condition while maintaining a set of all constraints that appear on each possible execution path. When the traversal encounters a \ttt{return} statement, we record the path constraints for the paths that terminate at that \ttt{return}, along with the return value expression. To prevent an exponential explosion of paths when collecting path constraints, we minimize the path set on each traversal step to only include paths that increase overall branch coverage (see Section \ref{sec:approach:path_min}).

We provide a detailed description of the path constraint collection procedure in Appendix A in our supplemental materials. 

\subsubsection{Path Minimization}
\label{sec:approach:path_min}


\begin{figure}
\centering
\begin{subfigure}[b]{0.5\linewidth}
\begin{algorithmic}[1]
\fontsize{7}{9}\selectfont
\Procedure{minimizePaths}{paths}
\State minconstraints = \{\}
\State minpaths = \{\}
\For{path in paths}
    \State path\_constraints = splitConstraints(path)
    \If{any of path\_constraints is \tbf{not} in minconstraints}
        \State minconstraints = minconstraints $\cap$ path\_constraints
        \State minpaths = minpaths $\cap$ path
    \EndIf
\EndFor
\State \Return minpaths
\EndProcedure
\end{algorithmic}
        \caption{\label{alg:path_minimization}Path Minimization Algorithm.}
    \end{subfigure}
    \begin{subfigure}[b]{0.49\linewidth}
        \includegraphics[width=\textwidth]{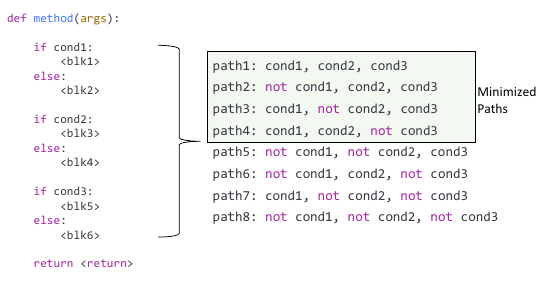}
        \caption{Example on three branches.\label{fig:path_min}}
    \end{subfigure}
\caption{Path minimization algorithmic definition and illustration of how path minimization prevents the number of paths from growing exponentially in the number of branches. A method with $n=3$ \ttt{if} conditions will have $2^n = 8$ possible execution paths, but applying the algorithm in \ref{alg:path_minimization} reduces the number of paths that need to be tested to at most $n+1 = 4$, each of which covers a unique branch condition.}
\label{fig:path_linearize_ex}
\end{figure}


One challenge in enumerating execution paths in a method is that the number of possible execution paths grows exponentially in the number of branches. Therefore collecting path constraints can result in a very large number of paths to test. This is undesirable because most of the paths will usually not add additional line or branch coverage and therefore are redundant from the perspective of a developer. Therefore, when collecting path constraints to construct path constraint prompts we only collect a \emph{linearly independent subset} of basis path constraints for use in prompting.

Basis paths were first proposed as a measure of method complexity~\cite{McCabe1976ACM} and are referred to a basis paths because they form a linear basis for all paths when expressed as a set of columns in the adjacency matrix of the method control-flow-graph. For testing purposes, basis paths are convenient because a set of tests that execute a set of basis paths in a method will achieve full branch coverage on that method.

Algorithm \ref{alg:path_minimization} describes how we compute basis paths from a set of path constraints. We first split each path into its constituent branch constraints, and check if any of the path's constraints are not in the set of linearized path constraints. If a path includes a branch constraint that is not in the linearized constraint set, we add it to the set of linearized paths and add its branch constraints to the set of linearized path constraints. 

Figure \ref{fig:path_linearize_ex} illustrates how path minimization reduces the number of collected paths for a simple example method with three sequential \ttt{if-else} branches. The total number of paths in the method is $2\times2\times2=8$, but applying Algorithm \ref{alg:path_minimization} reduces the number of paths to at most 4 (one initial path, $+1$ for each branch). We apply path minimization after collecting new path constraints from any \ttt{if} or \ttt{while} statement while traversing the focal method AST.



\subsection{Step-II.~Context Construction}

\begin{figure}
\centering
        \includegraphics[width=0.7\textwidth]{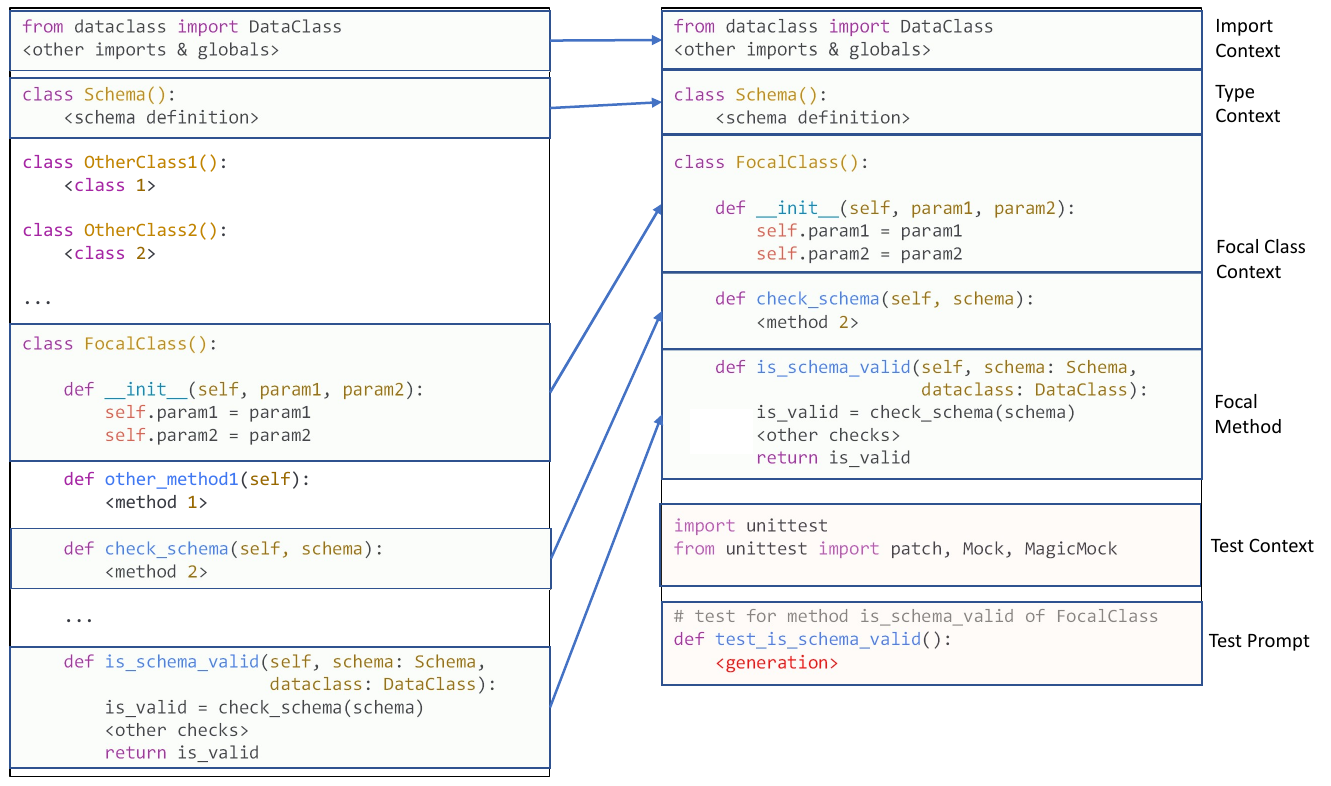}
\caption{Illustration of context construction on a focal method \ttt{is\_schema\_valid}. The context is composed of 5 components:
1. \emph{Imports and Globals.} Imported modules and classes that appear in the focal file, along with global variables defined in focal file.
2. \emph{Type Context.} Definitions of types that are used in the focal method and defined in the focal file.
3. \emph{Focal Class Type Context.} The definition of the focal class type signature and initialization.
4. \emph{Focal Class Method Context.} Definitions of any focal class methods that are called in the focal method.
5. \emph{Focal Method.} The definition of the focal method to be tested.
Constructing the generation context to expose relevant types and dependencies helps the model to attend to relevant definitions when generating. All objects that are included in the context are included in the test execution context and override any import statements generated by the model. This is particularly beneficial for test generations with GPT models, which we found are prone to generating hallucinated import statements (see Section \ref{sec:eval:gpt}).\label{fig:adaptive_context}}
\end{figure}

Prior work in test generation with LLMs have demonstrated that including additional context to the focal method definition is beneficial to the quality of test generations~\cite{tufano2020unit}.
For \approach, we focus on exposing two types of context to the model that are beneficial to generating correct test cases: (i) \emph{Type context.} Context that includes type definitions for focal method parameters guides the model to generate correct test inputs, particularly when the input involves complex data structures. (ii) \emph{Dependency Context.} Dependency context includes imports and other definitions that are used in the focal method, such as API calls and global data structures.
We construct the generation focal context selectively to include only type context and dependency context.
In addition, we construct the test execution context based on the focal context to override any model-generated imports and type definitions. We found that this prevents many common errors, particularly when generating tests with chat-tuned models such as GPT-4 (See Section \ref{sec:eval:gpt}).

Figure \ref{fig:adaptive_context} illustrates how we construct the generation context. We first parse the focal file and extract all import statements, global definitions, class definitions, and method definitions. We first add all import statements and global definitions to the context. We then check which classes and functions are used in the focal method and add their definitions to the context window if they are defined in the focal file. If the focal method is defined in a class, we extract the class's type signature and constructor definition, and then include the definitions of any methods in the focal class that called in the focal method. Finally, we append the full focal method definition at the end of the context window immediately before the test context and test generation prompt.


\subsection{Step-III.~Test Generation}



\begin{figure}
\centering
        \includegraphics[width=0.9\textwidth]{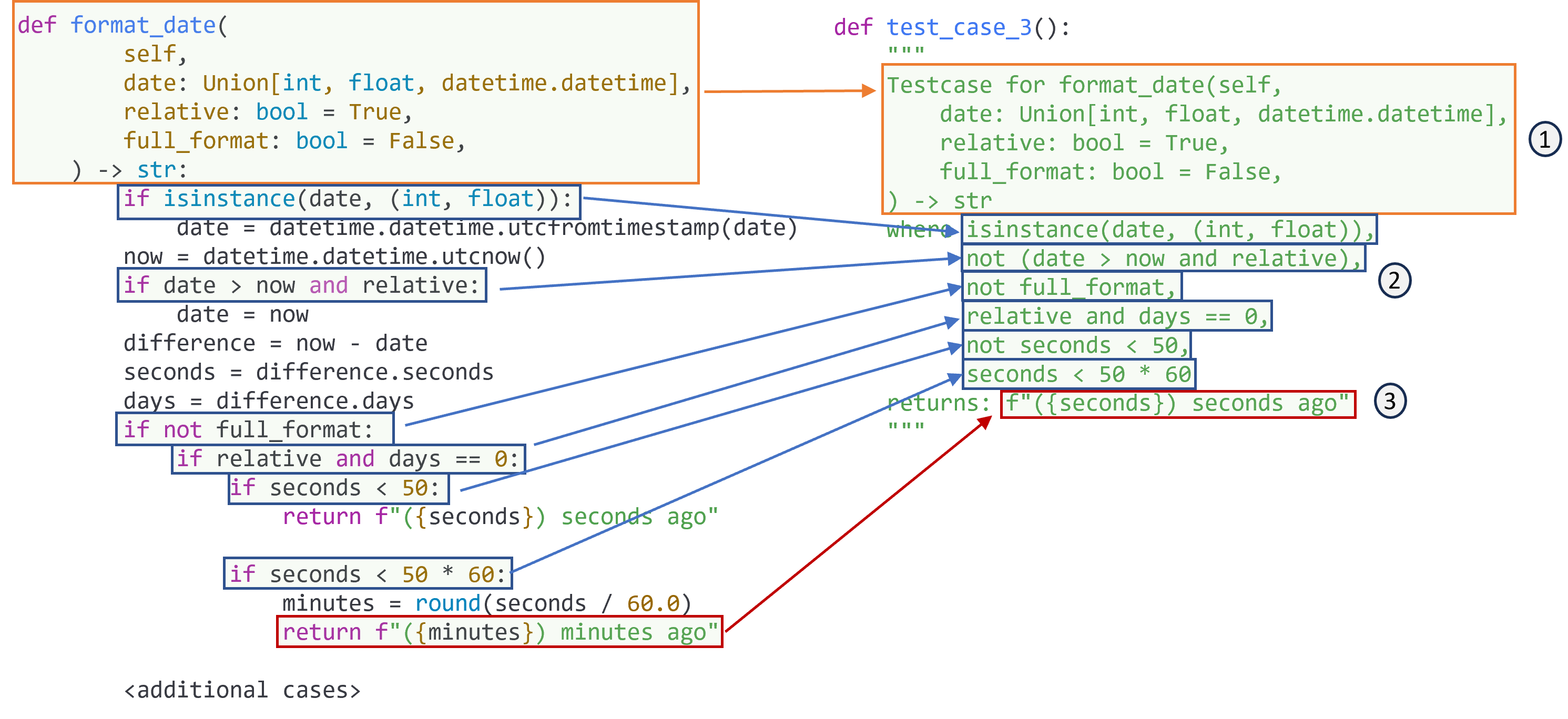}
\caption{Example of path constraint prompt construction on a simplified focal method in the \ttt{tornado} project. Each path prompt has three components: \circled{1} \tbf{Method Signature.} The prompt specifies the test case is for the focal method, including its full signature. \circled{2} \tbf{Path Constraints.} Next, the prompt specifies what path constraints should be satisfied by the inputs in the given testcase in order to follow the desired execution path. \circled{3} \tbf{Return Behavior.} The prompt specifies what return behavior, if any, is expected on the specified execution path. This can guide correct generation of assertions for specific return cases.}
\label{fig:prompt_construction}
\end{figure}

Once a set of path constraints have been collected for a given focal method, 
we construct prompts to generate tests for each path based on the focal method signature, path constraints, and return behavior on each collected execution path. Figure \ref{fig:prompt_construction} shows how a prompt is constructed for a single path in a simplified focal method for serializing a timestamp value. Each prompt first specifies the focal method the testcase is for along with the focal method signature, which serves to guide the model to correctly generate the focal method call in the test. The path constraints are combined to indicate that all constraints should be applied when generating the given test case. If the path used to generate the prompt terminates in a return statement, a \ttt{returns: <return\_value>} specifier is added to the prompt to guide assertion generation on the return value.



After constructing a set of path constraint prompts, we use these to iteratively prompt the model to generate tests for each path, and include the generated tests in the prompt for the next generation. This iterative prompting procedure, illustrated in Figure \ref{fig:end2end}, reduces the challenging problem of generating a high coverage testsuite into a multistage reasoning procedure based on individual execution paths and test inputs. Following common practice in code generation, for each generation, we check if the output code can be parsed, and if there are parse errors delete lines from the end of the generated code until it parses without errors. This eliminates errors caused by truncated model outputs.

Once a full set of tests have been generated, we construct the test execution context by importing all defined classes and variables that appear in the generation context. In addition, we copy all of the import statements that appear in the generation context and include them in the execution context. If the model generated any import statements for objects that are imported in the execution context, we remove the model-generated import. 

%% file: sections/05_evaluation.tex
\section{Evaluation}
\label{sec:eval}

We address the following research questions in our evaluation:
\begin{itemize}
    \item \tbf{RQ1.~Performance Impact.} How does \approach affect testing performance over simple test generation prompting methods in a regression setting?
    \item \tbf{RQ2.~Training Data Memorization.} How does \approach perform on projects that do not appear in the model training data?
    \item \tbf{RQ3.~Design Choices.} How do path constraints and calling contexts each contribute to the performance gains achieved by \approach?
     \item \tbf{RQ4.~Performance Impact on Large Models.} Does \approach still improve performance 
     on very large models?
\end{itemize}

\parheader{Experiment Setting.} We perform all evaluations on an AWS p4d.24xl instance with 8 Nvidia A100 GPUs and 96 vCPUs. We use Python 10 and Pytorch 1.13 with the Huggingface transformers framework to run local models.


\parheader{Evaluation Metrics.} Based on prior work on test generation~\cite{tufano2020unit, schafer2023adaptive}, we use the following metrics in evaluation:
\begin{enumerate}
    \item \tbf{Pass@1:} The average number of tests that run without errors and pass in each generated testsuite for each focal method when executed.
    \item \tbf{FM Call@1:} The average number of tests that correctly call the focal method in each generated testsuite for each focal method.
    \item \tbf{Correct@1:} The average number of tests that both pass and correctly call the focal method for each generated testsuite.
    \item \tbf{Line \& Branch Coverage:} Average line and branch coverage on the focal method for each testsuite generation.
\end{enumerate}
These metrics are computed similarly to the standard Pass@1 metric used in program generation benchmarks such as HumanEval~\cite{chen2021evaluating} and MBPP~\cite{austin2021program}, but may contain partially passing rates on each generation, since a single generated testsuite may contain both passing and failing tests.


\parheader{Benchmark Programs.} We evaluate on \evalfms focal methods drawn from 26 open source projects used in benchmarks in prior work~\cite{bugsinpy,lemieux2023codamosa}. We select focal methods from these projects where Pynguin~\cite{lukasczyk2022pynguin} (an SBST tool for Python) was unable to achieve full coverage on the focal method during 10 runs, indicating the focal method poses a challenge for existing automated test generation tools. In this dataset, Pynguin had an average line coverage of 72.4\% with std. deviation of 12.7\%.


\parheader{Evaluated Models.} Following the literature on test generation with language models~\cite{tufano2020unit,alagarsamy2023a3test}, we evaluate \approach with a single recent open source model, CodeGen2~\cite{nijkamp2023codegen2}. To measure \approach's ability to generalize to larger closed source models, we additionally perform an evaluation with GPT-4~\cite{openai2023gpt4}.

\subsection{RQ1: Performance Improvement}
\label{sec:eval:perf_improvement}

We first evaluate how \approach affects testing performance over simple test generation prompting methods in a regression setting.

\parheader{Evaluation.} We evaluate RQ1 on CodeGen2~\cite{nijkamp2023codegen2}, a recently released open source code model with 16 Billion parameters and a mixed causal training objective with span corruption and infilling. 
We use two baselines: the pynguin-generated test cases, and
and a test completion prompt based on prior work using Codex for test generation~\cite{lemieux2023codamosa}. 
For each focal method and prompting strategy, we performed 10 generations with CodeGen2 and averaged the results for each focal method over the 10 generations.

\begin{figure}
\centering
    \begin{subfigure}{0.20\linewidth}
    \centering
        \includegraphics[width=\textwidth]{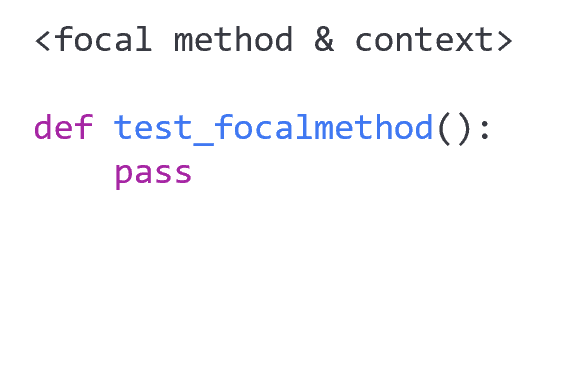}
        \caption{No-Op test.\label{fig:baseline1}}
    \end{subfigure}
  \begin{subfigure}{0.30\linewidth}
  \centering
        \includegraphics[width=\textwidth]{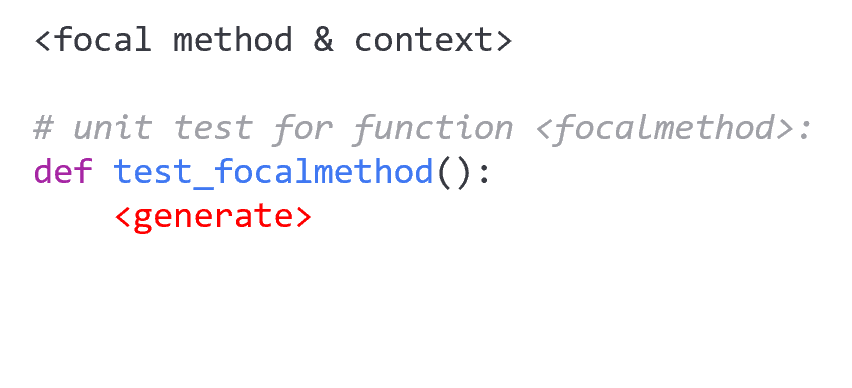}
        \caption{Baseline test prompt.\label{fig:baseline2}}
    \end{subfigure}
    \begin{subfigure}{0.46\linewidth}
  \centering
        \includegraphics[width=\textwidth]{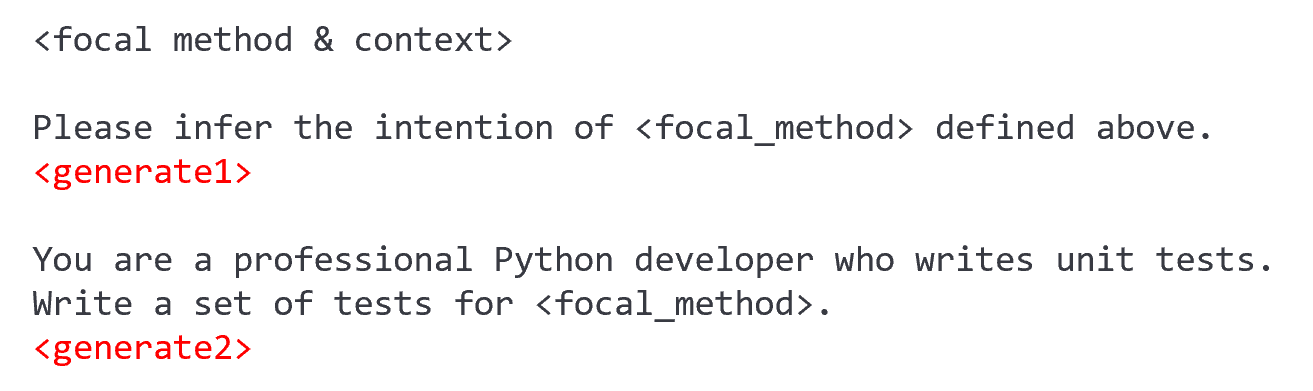}
        \caption{GPT Describe-Generate prompt.\label{fig:baseline3}}
    \end{subfigure}
\caption{Baseline prompts used in evaluation. 
\label{fig:baselines}}
\end{figure}

We also include as a baseline No-Op tests which simply load the target module in our execution framework and then return without explicitly running any test cases. In many cases loading the module containing the focal method will cause a significant proportion of the focal method to be covered, since both the method signature and docstring are counted as covered on load. The No-Op test format is shown in Figure \ref{fig:baseline1}. 
The test completion prompt is formatted as a partial test function that the model fills in. First a comment specifies the method is a test of the focal method, followed by the function signature for the a test of the focal method, as shown in Figure \ref{fig:baseline2}. 

\parheader{Observations.} Table \ref{tab:main_eval_results} summarizes our results for the evaluation of RQ1. Overall, we found that \approach significantly improves performance both in generating tests that execute with/without errors and call the focal method, and also significantly improve coverage over the model tests generated with the baseline prompt. We found the \approach is beneficial in two ways: first, the structure of path constraint prompts guides the model to generate tests by placing relevant information about how to call the focal method correctly based on its signature and how to generate correct assertions based on the expected method return behavior. This contributes to tests generated with \approach improving pass rate by a factor of nearly $4\times$  and correct rate, where the test both passes and calls the focal method, by a factor $5\times$. We found that while the model could often generate focal method calls 34\% of tests on average with the baseline prompt, most of these generations had incorrect focal method usage or other errors that prevented them passing, resulting in an overall correct generation of only 3\% on average.

The tests generated by \approach also achieve a 10\% improvement in line coverage and 4\% in branch coverage over the baseline prompts, indicating that path constraint prompts are also effective at guiding the model to generate higher coverage tests that exercise more distinct use cases for the target method. Moreover, if the line coverage that occurs on module load is taken into account by measuring improvement over No-Op tests, then the baseline prompts only improve absolute coverage by 5\% while \approach improves coverage by 15\%, a $3\times$ improvement.

On average, Pynguin achieves 72\% line coverage and 64\% branch coverage this benchmark, which is still significantly higher than the coverage of the tests generated by CodeGen2 with SymPrompt. As has been observed in prior work, test generation model results are biased by errors in model generations that cause many test suites to fail before they can execute the focal method~\cite{schafer2023empirical}. In contrast, during Pynguin's mutation-based testing process, it randomly generates and executes many different test candidates, and mutations that fail to improve coverage \eg{to errors} are discarded. Therefore, we also run a version of SymPrompt (Symprompt Filtered) where test suites with no working tests \ie{that do not execute and call the focal method} are discarded. Under this setting, CodeGen2 SymPrompt compares favorably with Pynguin (77\% line coverage, 66\% branch coverage on average).

\begin{figure}
\centering
    \begin{subfigure}[b]{0.412\linewidth}
    \centering
        \includegraphics[width=\textwidth]{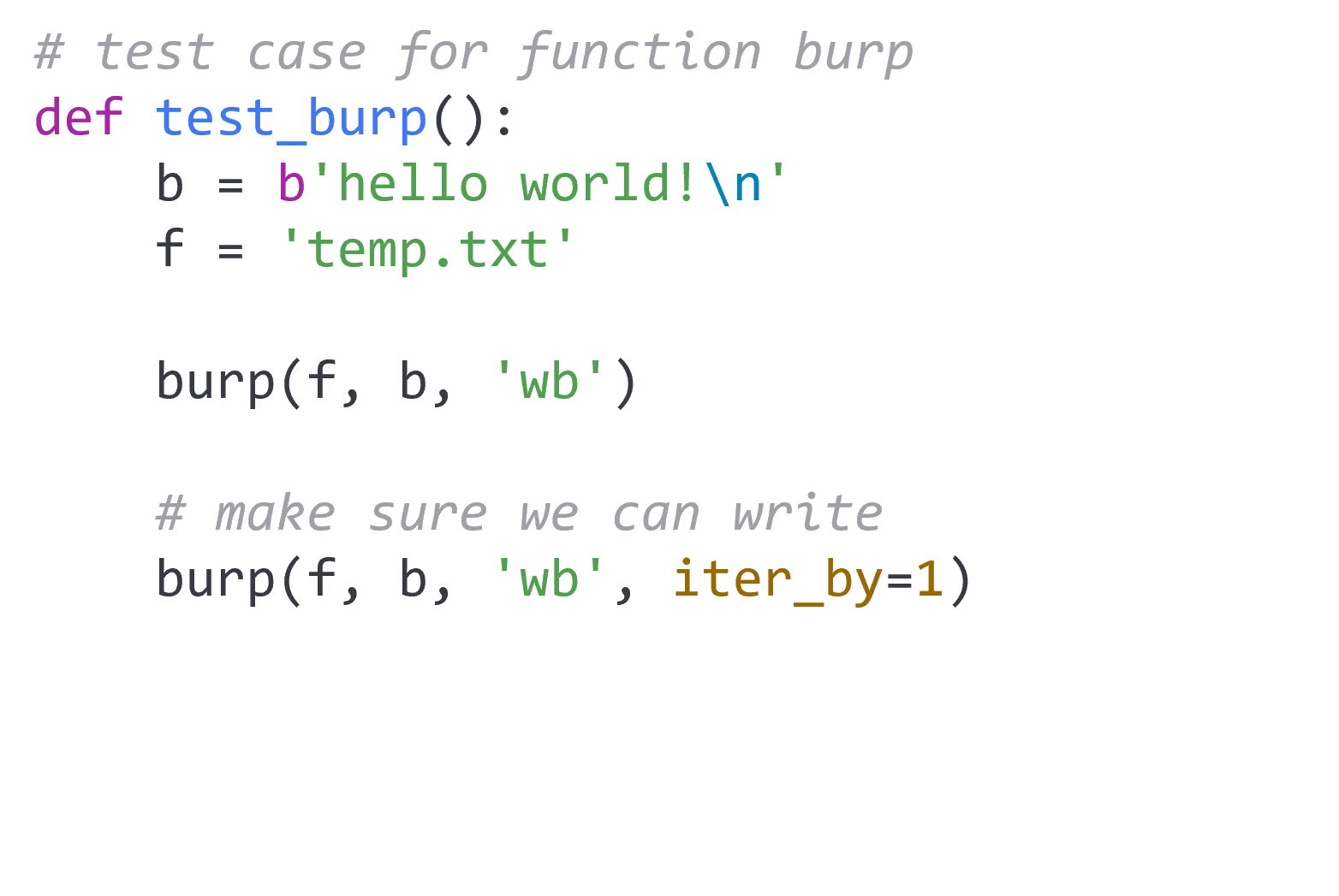}
        \caption{Baseline test generation.\label{fig:casestudy1:baseline}}
    \end{subfigure}
  \begin{subfigure}[b]{0.56\linewidth}
  \centering
        \includegraphics[width=\textwidth]{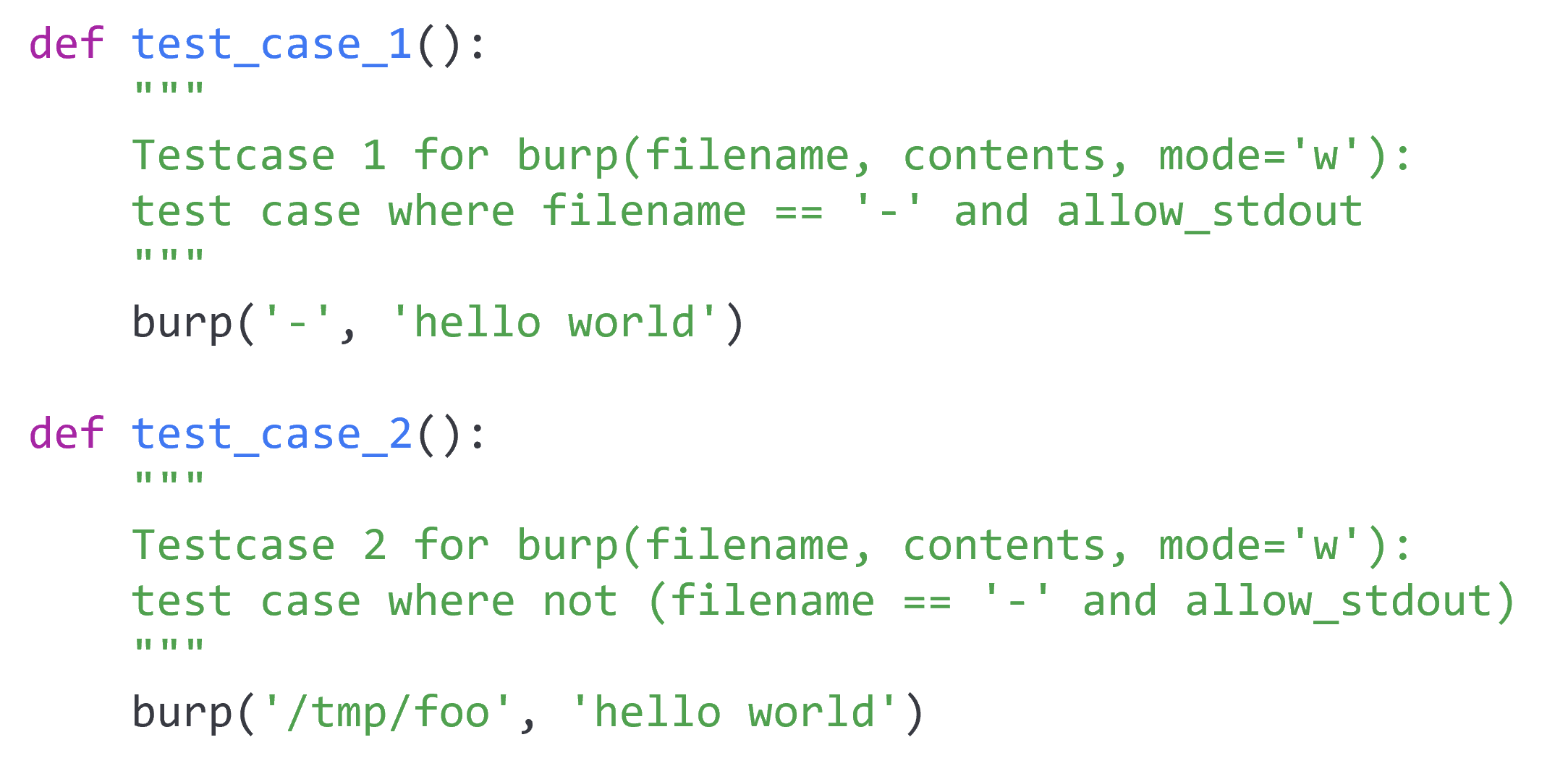}
        \caption{\approach test generation.\label{fig:casestudy1:symprompt}}
    \end{subfigure}
\caption{Case study of a simple focal method \ttt{burp} in the \ttt{pytutils} project with two main execution paths, based on whether the input filename is \ttt{-} or not. A test generation using the baseline prompt executes the method twice, but does not check for \ttt{filename='-'} as a test input. Moreover, the send focal method call uses a nonexistent parameter, \ttt{iter\_by}, preventing the test from fully executing. In contrast, \approach tests both paths and uses method correctly. Note that in this case the method does not return a value, therefore \approach does not prompt, {\em correctly}, for assertions on the return statement.\label{fig:casestudy1}}
\end{figure}

\parheader{Case Studies.} We found that \approach improved test generation in two ways: it reduced errors in calling the focal method by giving more precise guidance, and it helped to generate higher coverage test cases by testing more paths. The case study shown in Figure \ref{fig:casestudy1} illustrates both of these cases for a focal method \ttt{burp} from the \ttt{pytutils} project. The test shown in Figure \ref{fig:casestudy1:baseline} only tests one of the two execution paths in the \ttt{burp} method, where the input parameter \ttt{filename} is set to a normal value, and has an error where the focal method is called with an incorrect parameter. In contrast, the \approach generated tests call \ttt{burp} correctly and use test inputs to cover both the regular and \ttt{filename=`-'} paths, even though \ttt{`-'} is not a natural name for a file.

\begin{figure}
\centering
\sbox{\measurebox}{%
  \begin{minipage}[b]{.40\textwidth}
    \begin{subfigure}[b]{\linewidth}
        \includegraphics[width=\textwidth]{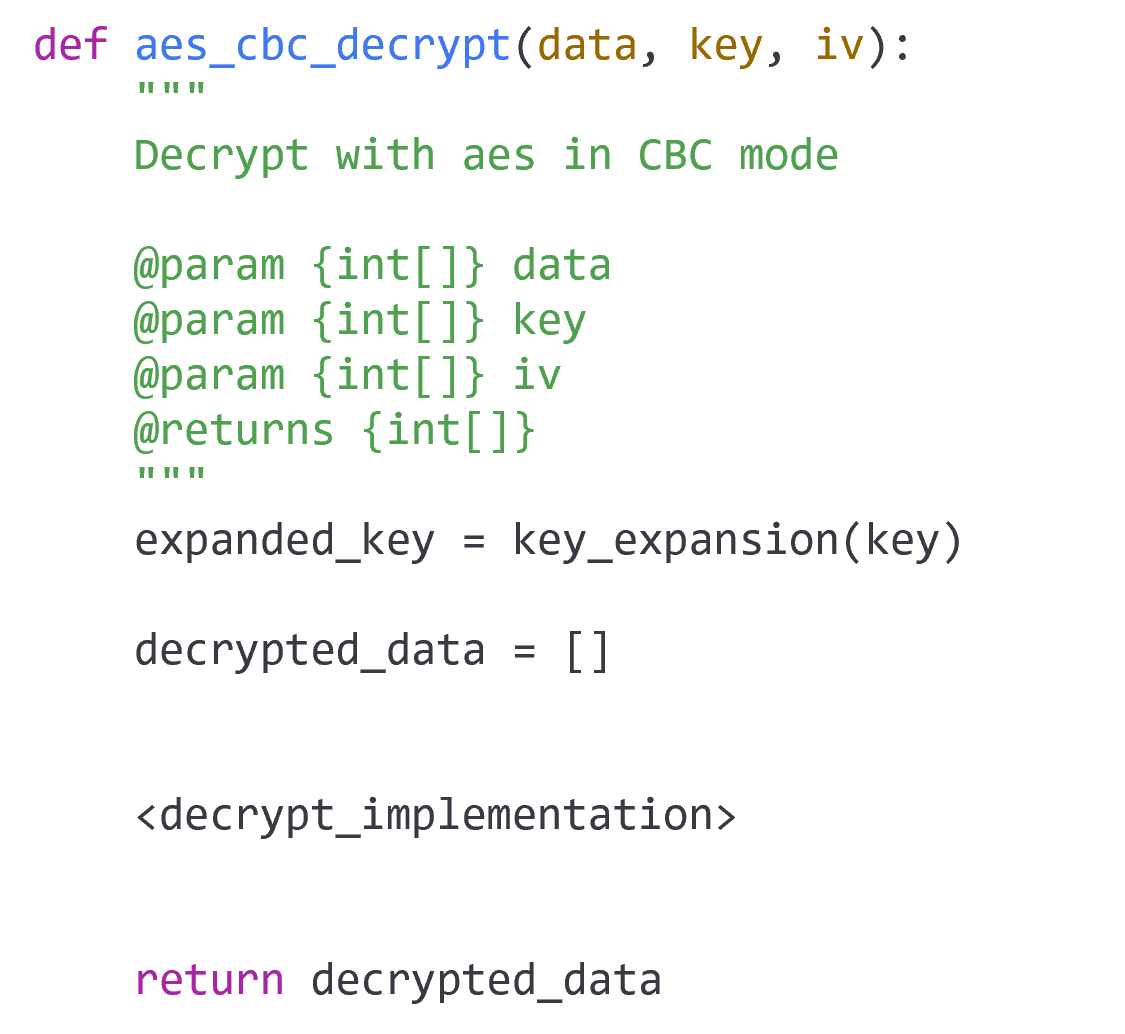}
        \vspace{-10pt}
        \caption{Focal method.\label{fig:casestudy_sbst_llm:fm}}
    \end{subfigure}
  \end{minipage}
}
\usebox{\measurebox}\qquad
\begin{minipage}[b][\ht\measurebox][s]{.46\textwidth}
\centering
      \begin{subfigure}[b]{\linewidth}
        \includegraphics[width=\textwidth]{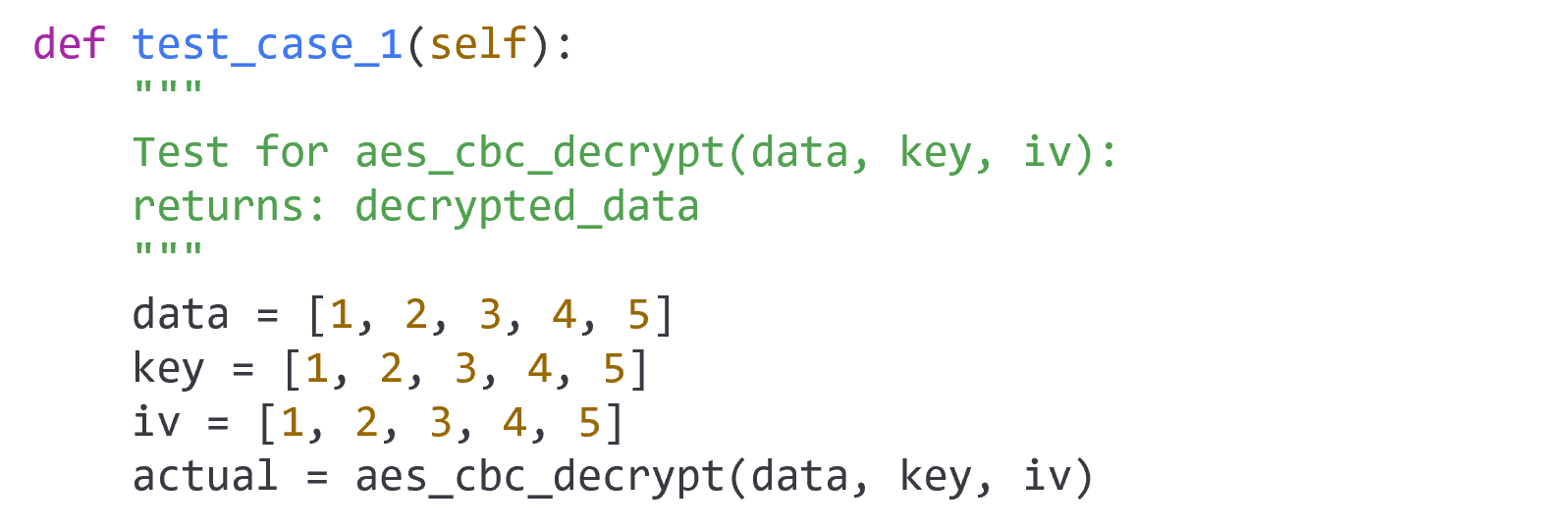}
        \vspace{-10pt}
        \caption{\approach test generation.\label{fig:casestudy_sbst_llm:llm}}
    \end{subfigure}
    
\vfill

  \begin{subfigure}[b]{\linewidth}
        \includegraphics[width=\textwidth]{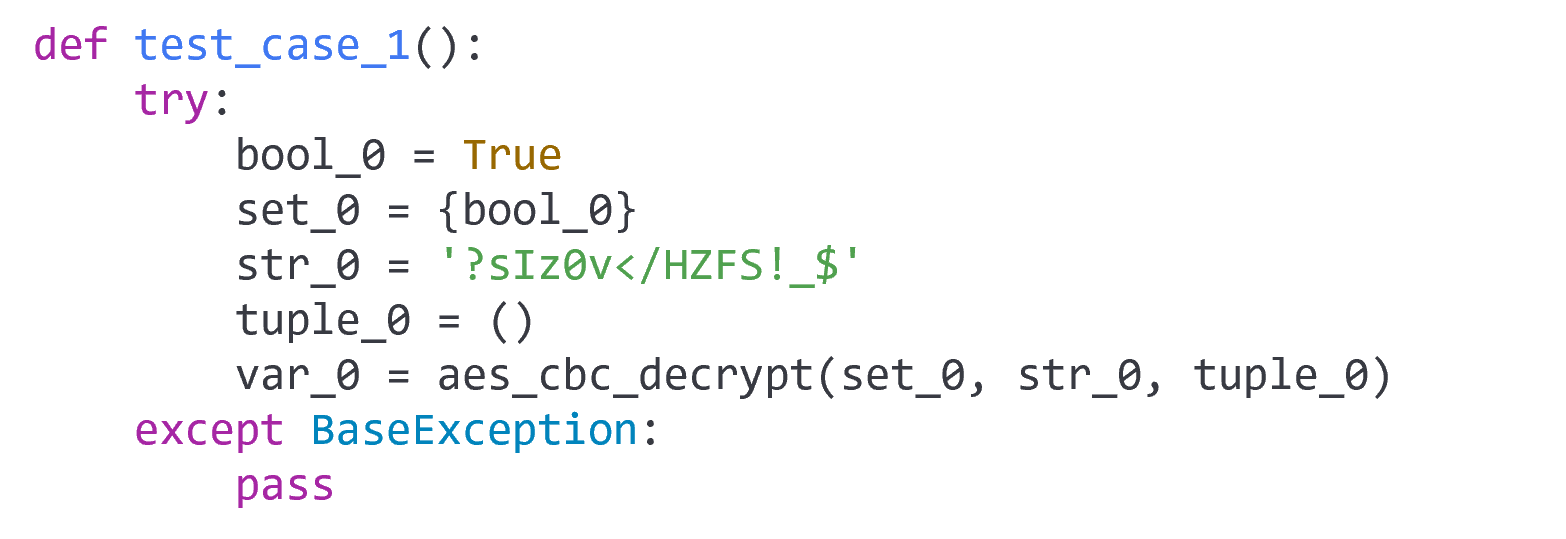}
        \vspace{-10pt}
        \caption{SBST test generation.\label{fig:casestudy_sbst_llm:sbst}}
    \end{subfigure}
\end{minipage}
\caption{Example test generations from an SBST tool (Pynguin) and LLM (CodeGen2) for focal method \ttt{aes\_cbc\_decrypt}. The method operates on arrays of ints, but an SBST approach is unable to infer types (even if they are specified in the comments, and generates test inputs that immediately raise an exception on the \ttt{key\_expansion(key)} call. An LLM can infer input types from context and comments and therefore generate a test case with correctly typed inputs that execute the entire method without errors.\label{fig:casestudy_sbst_llm}}
\end{figure}

Figure \ref{fig:casestudy_sbst_llm} illustrates how \approach can benefit testing a method that is challenging for an SBST approach due to specific type requirements for its input parameters. The input types are specified in a comment that an SBST tool cannot leverage, and as a result it generates test inputs that cause an exception on the first line of the method (see Figure \ref{fig:casestudy_sbst_llm:llm}). A language model however can use the comments for type hints along with the method signature to generate a correct method call that executes without errors (see Figure \ref{fig:casestudy_sbst_llm:llm}).

\begin{table}
\caption{\tbf{RQ 1. Results (LHS):} Results for evaluation of CodeGen2 test generation on \evalfms hard-to-test focal methods are shown in the left 4 columns. \approach is effective at guiding the model to call the focal method correctly and generate passing tests, as well as covering a wider range of use cases in its generated tests. These result in relative improvements of $5\times$ more correct test generations (which call the focal method and pass) and a 10\% improvement in coverage over tests generated with a baseline test completion prompt. \approach Filtered shows results when test generations with errors are discarded to provide a controlled comparison with Pynguin. Under this setting, \approach compares favorably with Pynguin, achieving marginally better line and branch coverage. \tbf{RQ 2. Results (RHS):} Results for evaluation on focal methods unseen in training data (based on AmIInTheStack tool~\cite{inthestack}) are shown in the right 4 columns. Compared to
the full benchmark results, both the baseline prompts and \approach have lower rates of correct test generations and coverage. However, tests generated with \approach still have significantly better correct generation rates ($4\times$ improvement) and coverage over the baseline prompt generated tests.\label{tab:main_eval_results}}
\centering
\fontsize{8}{9}\selectfont 
\setlength{\tabcolsep}{0.25em}
\begin{tabular}{
    >{\raggedright\arraybackslash}p{2.3cm} 
    R{1.0cm} R{1.0cm} R{1.0cm} R{1.0cm} R{1.0cm}|R{1.0cm} R{1.0cm} R{1.0cm} R{1.0cm} R{1.0cm}
}
\toprule
 & \multicolumn{5}{c|}{Full Benchmark} & \multicolumn{5}{c}{Unseen Projects} \\
Method & Pass@1 & FM Call@1 &  Correct@1 & Line Cov. & Branch Cov. & Pass@1 & FM Call@1 & Correct@1 & Line Cov. & Branch Cov. \\
\midrule
No-Op Tests & 1.00 & 0.00 & 0.00 & 0.33 & 0.33 & 1.00 & 0.00 & 0.00 & 0.26 & 0.36 \\
Baseline Prompt & 0.12 & 0.34 & 0.03 & 0.38 & 0.40 & 0.12 & 0.32 & 0.03 & 0.32 & 0.45 \\
\approach & 0.41 & 0.49 & 0.15 & 0.48 & 0.44 & 0.41 & 0.35 & 0.12 & 0.36 & 0.35 \\ \hline
Pynguin & - & - & - & 0.72 & 0.64 & - & - & - & 0.68 & 0.57 \\ 
\approach Filtered & 0.81 & 1.00 & 0.81 & 0.77 & 0.66 & 0.83 & 1.00 & 0.83 & 0.71 & 0.57 \\
\bottomrule
\end{tabular}
\end{table}

\subsection{RQ2: Training Data Memorization}
\label{sec:eval:data_memorization}

Since training data memorization can bias results with language model evaluations, we evaluate performance separately on a subset of projects that are not included in CodeGen2's training data.

\parheader{Evaluation.} To evaluate the potential impact of training data memorization on the results shown in RQ1, 
we conduct a seperate performance evaluation exclusively on projects that were excluded from CodeGen2's training data. Since CodeGen2 is trained on a subset of the Stack~\cite{kocetkov2022stack}, we use the AmIInTheStack tool~\cite{inthestack} to identify three projects in the evaluation set that not included in the stack and evaluate on the focal methods drawn from these projects. 

\parheader{Observations.} Table 
\ref{tab:main_eval_results}
shows results of this evaluation. Compared to the large scale evaluation results, CodeGen2's test generations with both baseline prompts and \approach have lower rates of correct test generations and coverage. However, the tests generated with \approach still have significantly higher rates of correct test generations ($\tfrac{.12}{.3} = 4\times$) and has 12.5\% higher line coverage relative to the LLM baseline. These results indicate that \approach is beneficial for generating more correct tests with higher coverage when focal methods are different from those seen in the training data. 

\subsection{RQ3: Design Choices}
\label{sec:eval:ablation}

\begin{table}
\caption{\tbf{RQ 3. Results (LHS):} Ablation results evaluated on 100 randomly sampled focal methods from the benchmark used in RQ1. The ablation results demonstrate that both type and dependency in calling context and path constraint prompts improve correct generations and coverage relative to baseline prompts, but path constraint prompts contribute significantly more to the overall improvement in correct generations and coverage demonstrated by \approach for generations with CodeGen2. \tbf{RQ 4. Results (RHS):} Results with GPT. When evaluating GPT we use as a baseline a two stage prompt in which the model is first prompted to describe the method under test, and then generate tests based on both the method definition and the previously generated description. Although performance is similar for both prompting approaches without calling context, when \approach is used without ablation it gives a relative improvement of 178\% in Correct@1 rate and 1.05\% relative improvement in coverage over baseline prompts.\label{tab:ablation_results}}
\centering
\fontsize{8}{9}\selectfont 
\setlength{\tabcolsep}{0.25em}
\resizebox{\textwidth}{!}{
\begin{tabular}{
    >{\raggedright\arraybackslash}p{2.0cm} 
    R{1.0cm} R{1.0cm} R{1.0cm} R{1.0cm} R{1.0cm}|R{1.0cm} R{1.0cm} R{1.0cm} R{1.0cm} R{1.0cm}
}
\toprule
 & \multicolumn{5}{c|}{CodeGen2} & \multicolumn{5}{c}{GPT-4} \\
Method & Pass@1 & FM Call@1 & Correct@1 & Line Cov. & Branch Cov. & Pass@1 & FM Call@1 & Correct@1 & Line Cov. & Branch Cov. \\
\midrule
Baseline Prompt & 0.09 & 0.37 & 0.04 & 0.30 & 0.29 & 0.14 & 0.12 & 0.09 & 0.36 & 0.40 \\
Constraints Only & 0.27 & 0.49 & 0.17 & 0.49 & 0.38 & 0.15 & 0.15 & 0.10 & 0.39 & 0.43 \\
Context Only & 0.16 & 0.40 & 0.06 & 0.42 & 0.35 & 0.38 & 0.28 & 0.18 & 0.43 & 0.47 \\
SymPrompt & 0.50 & 0.65 & 0.26 & 0.53 & 0.42 & 0.46 & 0.39 & 0.25 & 0.74 & 0.74 \\
\bottomrule
\end{tabular}
}
\end{table}

\approach uses both selective type and dependency focal context and path constraint prompts to improve performance over baseline test completion prompts. We evaluate the impact of each of these components on \approach's performance.

\parheader{Evaluation.} We conduct an ablation to evaluate how each of these methods contributes to performance improvements in isolation. When ablating focal context, we use the local context around the focal method based on prior work~\cite{lemieux2023codamosa}. We evaluate on 100 randomly sampled focal methods from the benchmark used in RQ1.

\parheader{Observations.} Table \ref{tab:ablation_results} (LHS) shows results of the ablations. \approach with no ablations achieves an overall 26\% correct generation rate and 53\% coverage on average. Ablating path constraint prompts but retaining calling context reduces the FM call rate and substantially reduces pass rate, indicating that the additional guidance from path constraint prompting is very beneficial for generating tests with correct focal method calls. The path constraint ablation also has significantly lower coverage, indicating that the path constraints serve to generate more thorough test cases. 

The ablation of calling context also reduces the pass rate and FM call rate of generated tests, although much less than ablating path constraint prompting, and still has significantly better performance than a full ablation of both methods. These results indicate that while test generations with path constraint prompts benefit from using calling context, the additional guidance and structure provided to the model in the symbolic prompts are crucial to the performance improvements in correct generations and coverage by \approach over baseline test generation prompts.

\subsection{RQ4: Large Model Performance Impact}
\label{sec:eval:gpt}




In addition to evaluating on an open source 16B parameter model, we also evaluate if the prompting strategy used by \approach can benefit test generations with a larger and better trained model. Recent work has shown that increasingly large scale language models exhibit \emph{emergent abilities} that are completely absent in smaller scale models~\cite{wei2022emergent}. In this evaluation we show that a significantly larger language model, GPT-4, exhibits the ability to reason precisely about path constraints in a focal method and generate its own path constraint prompts in a zero-shot setting. We find that prompting the model to approach test generation in this way is very beneficial for generating high coverage testsuites.


\parheader{Evaluation.} We evaluate with GPT-4~\cite{openai2023gpt4}, a significantly larger model than CodeGen2 that benefits from much more extensive training. 
We found that GPT-4 was capable of generating precise descriptions of the execution paths and constraints in focal methods given a 0-shot prompt, so instead of generating path prompts with static analysis, we prompt the model to describe execution paths and then embed each path description in a test docstring (See Figure \ref{fig:baseline3}).

Recent works have demonstrated that test generation with GPT models benefit from multi-stage prompts that incorporate description and planning~\cite{yuan2023no,li2023finding}, therefore as a baseline we use a 2 stage prompt that first asks the model to describe the intent of the method under test and then to generate a testsuite based on both the description and the focal context based on~\cite{yuan2023no}.

\parheader{Observations.} Table \ref{tab:ablation_results} (RHS) summarizes the results of our evaluation with GPT-4. We found that calling contexts in the generated tests were especially important for GPT-4s generations, since the model was prone to hullucinating incorrect import statements or using undefined classes and objects that were defined in the focal context. Overwriting the model-generated imports with the classes and objects identified in the calling context greatly reduced errors when the tests were executed. Using the path constraint prompts generated by GPT-4 in isolation did not lead to significant performance improvements over the baseline describe and generate prompts, but when used in conjunction with calling context the path constraint prompts improved the average coverage of generated tests over the baseline describe-generate prompt by a factor of more than 2, from 36\% to 74\%. We hypothesize that the significant performance difference occurs because path constraint prompts are effective for generating more high coverage testsuites, but most of those tests fail due to errors with imports and undefined variables when calling contexts are not used.

\begin{figure}
\centering
    \begin{subfigure}[b]{0.8\linewidth}
    \centering
        \includegraphics[width=\textwidth]{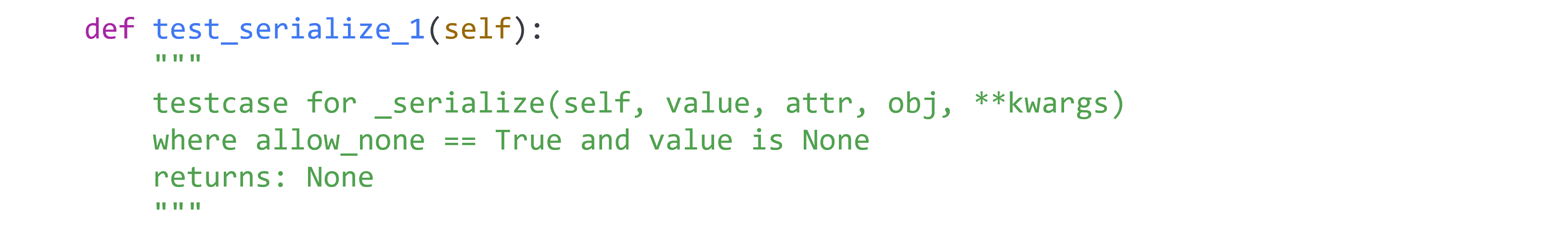}
        \caption{\approach path prompt.\label{fig:gpt_vs_symprompt:symprompt}}
    \end{subfigure}
    ~
  \begin{subfigure}[b]{0.8\linewidth}
  \centering
        \includegraphics[width=\textwidth]{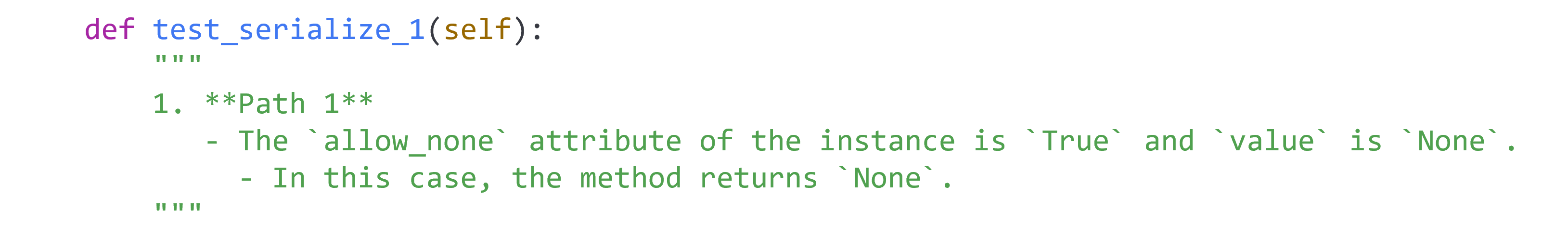}
        \caption{GPT-4 path prompt.\label{fig:gpt_vs_symprompt:gpt}}
    \end{subfigure}
\caption{Comparison of \approach-generated path prompt to GPT-4 generated path prompt. GPT-4 is capable of generating precise execution path descriptions with more natural language. We construct path prompts by using a markdown parser to extract each path description and embed them as docstrings for each test generation.\label{fig:gpt_vs_symprompt}}
\end{figure}

\begin{figure}
\centering
    \begin{subfigure}[b]{0.8\linewidth}
    \centering
        \includegraphics[width=\textwidth]{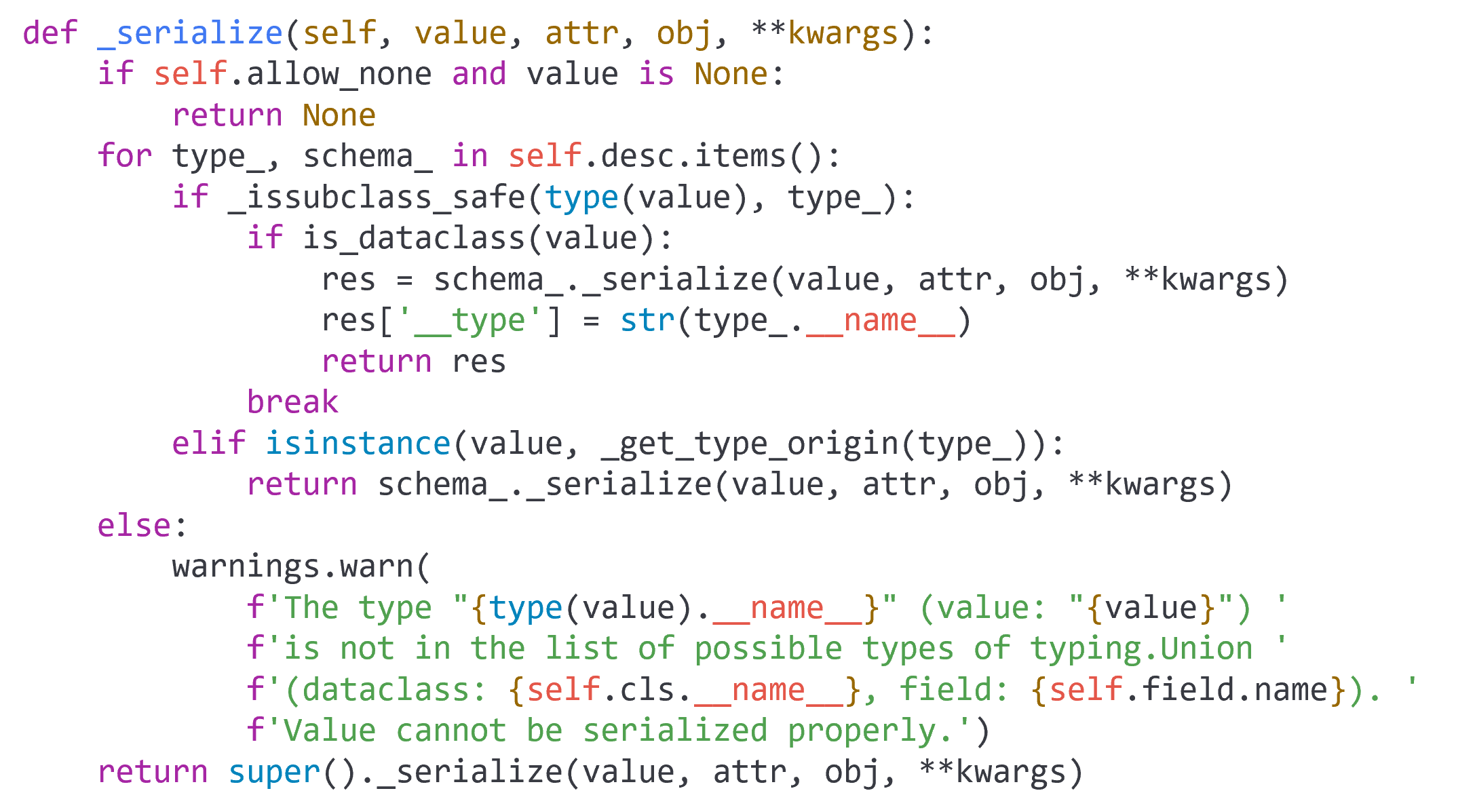}
        \caption{Focal method.\label{fig:gpt_casestudy:fm}}
    \end{subfigure}
  \begin{subfigure}[b]{0.95\linewidth}
  \centering
        \includegraphics[width=\textwidth]{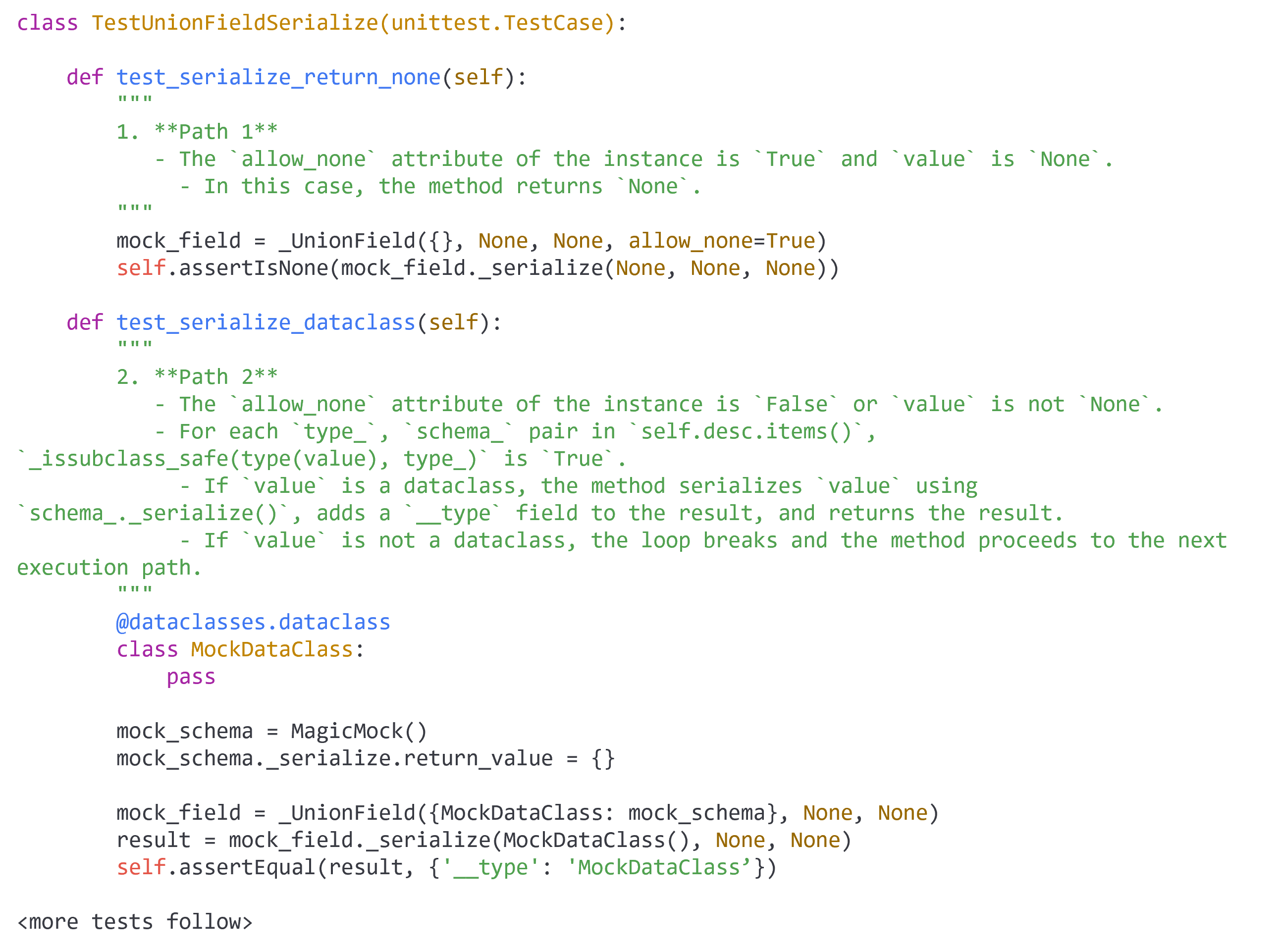}
        \caption{GPT-4 generated tests.\label{fig:gpt_casestudy:symgen}}
    \end{subfigure}
\caption{Case study showing GPT generations with path constraint prompts.\label{fig:gpt_case_study}}
\end{figure}

\parheader{GPT vs. Analysis-Generated Path Prompts.} Figure \ref{fig:gpt_vs_symprompt} shows a comparison of the path descriptions generated by GPT-4 and \approach's static analysis on the method \ttt{\_serialize} shown in Figure \ref{fig:gpt_casestudy:fm}. We found that, compared to the static analysis, GPT-4 was able to generate more natural path constraints while still giving precise descriptions. Figure \ref{fig:gpt_case_study} shows the test cases generated for two GPT path prompts on the method \ttt{\_serialize}.

\parheader{Path-Following Generation Accuracy.} In addition to measuring the impact of \approach on overall coverage, we conducted a small study of how effective the models are at generating tests that follow the specific paths specified in the prompts by manually examining 10 generated tests in the GPT-generated set. Of these, 6 out of 10 followed their specified paths. The missing four cases are either due to deeply nested branching or exception handling--the model either did not generate correct preconditions for deeply nested branches or error-handling paths. We also observed qualitatively that CodeGen2 usually generates test inputs that follow specified paths when the constraint involves an input parameter, as shown in Figure \ref{fig:working_ex}, while GPT-4 is also able to generate test inputs for more complex path constraints involving class variables and external function calls, as shown in Figure \ref{fig:gpt_case_study}.

%% file: sections/06_threats_to_validity.tex

\section{Threats to Validity \& Discussion}
\label{sec:threats}

\parheader{Model and Benchmark Validity:} Our evaluations focus on open source Python projects and utilize specific language models (CodeGen2 and GPT-4). This restricts the generalizability of our findings to other languages or models. However, these are two large state-of-the-art models. 
Also, none of the methods are specific to Python. 
So we expect the findings will hold consistently in other settings.

\parheader{Memorization Validity:} Although we use the AmIInTheStack tool to prevent training data memorization from biasing our RQ 2 results, the possibility remains that CodeGen2 could have seen some of the focal methods or similar code in its training data. However, since these models are not explicitly trained for test generation tasks, we think this threat is minimal.  

\parheader{Metric Validity:} Our evaluations are based on the metrics Pass@1, FM Call@1, Correct@1, and Line Coverage. However, these metrics might not capture the full complexity or usefulness of a generated test case.

\parheader{Test Generation in a Regression Setting:} In this paper, we operated under the assumption that the tests are generated within a regression setting, assuming the correctness of the underlying focal method implementation. As a result, our generated tests may not uncover any implementation bugs. This approach to testing also has limitations, particularly when the implementation of the focal method is not yet finalized, a scenario commonly encountered in continuous development environments.

%% file: sections/07_discussion.tex






%% file: sections/08_related_work.tex
\section{Related Work}


Our work relates to the following areas: Search Based Software Testing, Symbolic Test Generation, Test Generation with LLMs, and Hybrid LLM-SBST test generation.

\parheader{Search Based Software Testing (SBST) and Symbolic Approaches.}
Evosuite is an SBST regression testing framework for java that generates regression tests based mutation testing and coverage guided randomized test generation~\cite{fraser2011evosuite, fraser2013evosuite, fraser20151600}.
Randoop uses coverage guided randomized test generation for Java in conjunction with sanity checking oracles to check for common classes of bugs like NullException errors~\cite{pacheco2007randoop, pacheco2007feedback}.
Pynguin applies SBST to generate regression tests for Python~\cite{lukasczyk2020automated, lukasczyk2022pynguin}.
PeX is a whitebox test generation with concolic execution and constraint solver~\cite{tillmann2008pex}.
Korat tests based on a formal specification~\cite{boyapati2002korat}.
Dart performs concolic test generation~\cite{godefroid2005dart}.
Cute performs concolic testing for c~\cite{sen2005cute}.
Our approach is conceptually related to coverage driven SBST approaches and Concolic Execution because it formulates test generation as a constraint solving problem for the LLM, where the LLM is guided to generate a test that will follow a specific execution path. However, instead of performing symbolic constraint solving to follow specific execution paths we give the LLM access to the focal method source code.

\parheader{LLM Regression Test generation}.
Athenatest finetune pretrained transformers on paired method-test data and show including more focal context in prompt leads to higher coverage test generations~\cite{tufano2020unit}.
Bariess et. al. compare Codex-generated tests to randoop based on a one-shot test example and show the codex-generated tests achieve better coverage(arxiv)~\cite{bareiss2022code}.
A3Test apply postprocessing to correct test naming errors in model generations using PLBart(arxiv)~\cite{alagarsamy2023a3test}.
Hashtroudi et al. show that models finetuned on existing project testsuites output higher quality test generations(arxiv)~\cite{hashtroudi2023automated}.
MuTAP uses mutation testing to guide LLM test generations towards tests that are more likely to detect bugs and show improved bug detection with Codex and Llama2 on Defects4j(arxiv)~\cite{dakhel2023effective}.
These approaches use fixed prompting strategies, our work focuses on developing code-aware prompts that guide the model to generate a high coverage set of tests.

\parheader{ChatGPT test generation}.
Testpilot generates javascript tests with ChatGPT 3.5 using a 0-shot test prompt and then iteratively adds additional code and documentation context if the generated tests fail(arxiv)~\cite{schafer2023adaptive}.
ChatUnitTest generates tests using adaptive focal context based on the maximum focal length and then applies both rules-based repair and self-debugging based on error messages when generated tests fail~\cite{xie2023chatunitest}.
ChatTester similarly generates tests based on the focal context and then prompts ChatGPT with error messages when generated tests fail(arxiv)~\cite{yuan2023no}.
These approaches all focus on constructing the focal context and then fixing ChatGPT's generations. In contrast our work focuses on developing prompts to guide the model to test each execution path in the focal method, improving testsuite coverage.

\parheader{Hybrid SBST-LLM regression test generation.}
Codamosa runs Pynguin, an SBST tool for python, and iteratively calls Codex to generate additional testcases for methods with low coverage~\cite{lemieux2023codamosa}. Our work focuses on improving LLM testsuite generations instead.

%% file: sections/09_conclusion.tex
\section{Conclusion}

This paper introduces \approach, a novel approach to test generation with LLMs by decomposing the test suite generation process into a structured, code-aware sequence of prompts. \approach significantly enhances the generation of comprehensive test suites with a recent open source Code LLM, CodeGen2, and achieves substantial improvements in the ratio of correct test generations and coverage. Moreover, we show that when given a specific instruction prompt to analyze execution path constraints, GPT-4 is capable of generating its own path constraint prompts, which improves the coverage of its generating testsuites by a factor of $2\times$ over prompting strategies from recent prior work.